\documentclass{aa}  

\usepackage{graphicx}
\usepackage{txfonts}
\usepackage[colorlinks=true,
     linkcolor=blue,
     filecolor=blue,
     citecolor = blue,      
     urlcolor=blue,]{hyperref}
\usepackage{graphicx}	% Including figure files
\usepackage{subcaption}
\usepackage{amsmath}	% Advanced maths commands
\usepackage{amssymb}	% Extra maths symbols
\usepackage{siunitx}
\usepackage[dvipsnames]{xcolor}
\usepackage{float}
\usepackage[normalem]{ulem}

\usepackage{tikz}

\newcommand{\panellabel}[4][0mm]{%
  \begin{tikzpicture}
    \node[anchor=south west,inner sep=0] (img) {\includegraphics[width=#2]{#3}};
    \node[anchor=north west,font=\bfseries,inner sep=2mm,xshift=#1]
         at (img.north west) {#4};
  \end{tikzpicture}%
}

\addto\extrasenglish{}
\addto\extrasenglish{}
\addto\extrasenglish{}
\addto\extrasenglish{}
\addto\extrasenglish{}
\addto\extrasenglish{}
\addto\extrasenglish{}
\addto\extrasenglish{}
\newcommand{\aref}[1]{\hyperref[#1]{Appendix~\ref{#1}}}

\usepackage[nameinlink,capitalize]{cleveref}

\usepackage{xstring}
\creflabelformat{equation}{#2#1#3}
\crefformat{equation}{#2Eq.~(#1)#3}
\DeclareRobustCommand{\pcref}[1]{%
  \IfSubStr{#1}{,}{(Eqs.~\labelcref{#1})}{\hyperref[#1]{(Eq.~\ref*{#1})}}}
\definecolor{darkgreen}{rgb}{0.13, 0.55, 0.13}

\makeatletter
\renewcommand*\aa@pageof{, page \thepage{} of \pageref*{LastPage}}
\makeatother

\begin{document}

   \title{Quantifying collision-driven mass loss in supermassive star formation: the role of stellar structure and accretion}

   % \subtitle{I. Overviewing the $\kappa$-mechanism}

   \author{P. A. Solar
          \inst{1},
          B. Reinoso \inst{2}, 
          D. R. G. Schleicher \inst{3} , 
          \and 
          R. Banerjee \inst{1}
          }

   \institute{Hamburger Sternwarte, Universit\"at Hamburg, Gojenbergsweg 112, 21029 Hamburg, Germany\\
              \email{paulo.solar.vera@uni-hamburg.de}%\\
              %\email{robi.banerjee@uni-hamburg.de}
         \and
             Department of Physics, Gustaf Hällströmin katu 2, FI-00014, University of Helsinki, Finland\\
             \email{bastian.reinoso@helsinki.fi}
         \and
             Dipartimento di Fisica, Sapienza Università di Roma, Piazzale Aldo Moro 5, 00185 Rome, Italy\\ \email{dominik.schleicher@uniroma1.it}
         \\
             }

   \date{Received xxxx; accepted xxxx}
   
\titlerunning{Impact of mass loss on the formation of a supermassive black hole seed}
\authorrunning{P. A. Solar et al.}

% \abstract{}{}{}{}{} 
% 5 {} token are mandatory
 
  \abstract
  % context heading (optional)
  % {} leave it empty if necessary  
   {Observations of high-redshift galaxies with JWST have renewed interest in scenarios where supermassive stars form via runaway stellar collisions in dense clusters, yet the impact of collision-driven mass loss on their growth remains uncertain. In this work, we perform a post-processing analysis of 3D hydrodynamical simulations of the formation of a supermassive star, applying an analytic mass-loss prescription to stellar collisions while exploring different assumptions for the internal stellar structure. We consider a polytropic main-sequence model, a semi-analytic accreting protostar model, and structures derived from stellar evolution calculations. We find that the cumulative mass-loss fraction depends sensitively on the adopted stellar structure, ranging from $\lesssim 10-25\%$ for more compact configurations to $\gtrsim 30-40\%$ for more extended protostellar models. The importance of mass loss further depends on the dynamical state of the system, including the ratio of stellar velocity dispersion to the stellar surface escape velocity. We find significant uncertainty depending on the prescription used. As a result, collision-driven mass loss could significantly limit the growth of the central object, at least in some cases. Overall, our results indicate that uncertainties in the internal structure of rapidly accreting protostars represent a major source of systematic uncertainty and must be better constrained to robustly assess the viability of runaway-collision pathways for forming massive black hole seeds.}

   \keywords{methods: numerical - stars: formation – stars: Population III - stars: mass loss - quasars: supermassive black holes - early Universe}

   \maketitle
%
%________________________________________________________________

\section{Introduction}

Measurements of chemical abundances in galaxies are key to understanding their formation and evolution. Elements heavier than hydrogen and helium (metals) trace star formation because they are produced through stellar life cycles \citep{maiolino_re_2019}. From the observed mass-metallicity relation, we know that more evolved galaxies are more chemically enriched, while galaxies at higher redshift tend to show lower metallicities.

Recent JWST/NIRSpec observations of very high-redshift galaxies have motivated a range of theoretical scenarios to explain unusually strong $\rm N/O$ and $\rm C/O$ ratios \citep[e.g.][]{bunker_jades_2023,cameron_nitrogen_2023,yanagisawa_strong_2024}. These include AGN, standard stellar populations with fine-tuned models, Wolf-Rayet stars and transient phenomena such as tidal disruption events \citep[e.g.][]{feltre_2016,Kochanek_2016,Hirschmann_2019,jiang_evidence_2021}. A particularly interesting class of models invokes runaway stellar collisions in dense clusters, which can form very massive or supermassive stars (VMSs/SMSs) before the first supernovae \citep[e.g.][]{Portegies_Zwart1999,Portegies_Zwart2002,Gurkan_2004,Katz_2015}. In these models, collision products can be chemically well mixed, bringing CNO-processed material to the surface. The possible metal-poor VMSs could retain most of their mass until collapse, while metal-enriched VMSs may drive strong winds that contribute to light-element enrichment \citep[e.g.][]{glebbeek2009,vink_theory_2022}.

In collision-driven formation scenarios, the same mergers that build up an SMS also eject mass from the system. This collision-induced mass loss can reduce the growth efficiency of the central object and may affect the expected enrichment yield (e.g. by limiting the final SMS mass and modifying the structure of merger products). These effects are therefore important when considering whether runaway-collision channels can produce SMSs massive enough to explain the abundances observed by JWST. Determining the SMS mass range that can explain the observed abundance patterns is a key point for constraining SMS evolutionary channels in the early Universe. \citet{nandal2025} showed that the $\rm N/O$ ratio of GS~3073 at redshift $z=5.55$ can be explained by primordial Population~III stars with masses in the range $10^3-10^4~\rm M_{\odot}$. They found that stars below $10^3~\rm M_{\odot}$ cannot produce $\rm N/O$ ratios above $-0.16$, while for $M>10^4~\rm M_{\odot}$ the upper limit is $0.19$.

Although SMSs are a promising model for explaining the abundances observed by JWST, their formation and evolution are still unclear. In the literature, the formation of SMSs in the early Universe have been explored through gas-dominated or star-dominated channels. In the purely gas-dominated channel, isothermal collapse driven by atomic cooling in halos with virial temperatures $T_{\rm vir} > 8000~\rm K$ can produce SMSs with masses of order $\sim 10^5~\rm M_{\odot}$. In this scenario, efficient $\rm H_2$ cooling is suppressed by a strong Lyman-Werner UV background, by gas dynamical heating, or by hybrid $\rm Ly\alpha/H_2$ cooling moderated by LW radiation \citep[e.g.][]{omukai2001ApJ,bromm2003ApJ,Spaans2006ApJ,Latif2013MNRAS,latif2015,wise_formation_2019,prole2024}. In a purely stellar formation channel, high central stellar densities trigger a runaway-collision regime in star clusters, forming an SMS \citep{Portegies_Zwart2002}. \citet{escala2021} proposed the formation of an SMBH in nuclear stellar clusters triggered by runaway stellar collisions because the collision timescale is shorter than the cluster age, leading to the formation of a massive black hole. Based on this work, \citet{vergara_global_2023} defined a critical mass over which a nuclear star cluster can form an SMBH. This relationship between cluster mass and black hole formation efficiency was compared with numerical simulations and observations of diverse stellar systems, finding a direct correlation and a limiting black hole efficiency of $\epsilon \sim 0.1$ when $M/M_{crit}\sim0.3$ \citep{vergara_efficiency_2024,Rantala2025a,Rantala2025,Liempi2025,Vergara2026}. Additionally, a combination of both channels is likely, since gas present in some star clusters may contribute to the formation of central massive objects through gas accretion and stellar collisions \citep[e.g.][]{boekholt_formation_2018,Tagawa2020b,das_2021,schleicher2022,schleicher_physical_2023,reinoso_formation_2023,saavedra-bastidas_gravitational_2024,reinoso_2025,PSolar2025}. Clarifying the mechanisms and processes capable of explaining the initial masses of these SMSs is essential to understanding their evolution to the masses that we observe today. 

Due to the challenge in estimating both the impact of stellar collisions and the associated mass-loss fraction, most studies of runaway-collision scenarios neglect this effect. However, collision-driven mass loss has been extensively investigated in other astrophysical contexts, particularly in studies of blue stragglers \citep[e.g.][]{lombardi_collisions_1996,lombardi_stellar_2002,sills_blue_2005,gaburov_mixing_2008,glebbeek_structure_2013}. In particular, \citet{lombardi_stellar_2002} developed a semi-analytic model to determine the structure of stellar collision products based on conservation laws. The prescription was calibrated using smoothed particle hydrodynamics (SPH) simulations of non-rotating low-mass main-sequence (MS) stars, reproducing the thermodynamic and chemical profiles of detailed hydrodynamical collision models with good accuracy, as well as the subsequent evolution of the merger remnants.

Despite its simplicity, the assumption of fully conservative stellar collisions remains widely used in N-body simulations, even though it is not necessarily physically motivated. However, some studies have already explored SMBH formation through runaway collisions considering collision-driven mass loss via semi-analytic prescriptions \citep[e.g.][]{alister_seguel_formation_2020,2023Rose,reinoso_2025,2026Rantala}. Usually, these prescriptions depend primarily on the mass ratio of the merging stars, or on simplified comparisons between the kinetic and binding energies of the collision. However, the density profile of an MS star differs markedly from that of a highly accreting SMS, whose inflated envelope may strongly affect the amount of mass ejected during stellar collisions. Although semi-analytic prescriptions cannot capture the full hydrodynamical evolution of stellar collisions, they provide a computationally efficient and physically motivated framework for estimating collision outcomes in runaway-collision scenarios. The prescription adopted in this work depends explicitly on the internal stellar structure of the colliding stars and therefore captures the dependence of collision-driven mass loss on stellar compactness. In the context of SMS formation, where rapidly accreting stars develop highly inflated and weakly bound envelopes, the internal structure is expected to play a key role in determining the collision outcome. We therefore explore the impact of collision-induced mass loss during SMS formation in collapsing primordial gas clouds containing multiple protostars, focusing on how different assumptions about the stellar structure affect the collision-driven mass loss and the final mass of the central massive object (CMO).

\section{Methods} \label{sec:method}
%\subsection{Physical Setup} \label{sec:simulationSetup} 

We investigate this through a post-processing analysis of our hydrodynamic simulations presented in \citet{PSolar2025}. These simulations explored different gas temperatures to study how the initial instability of the system affects the efficiency of SMBH seed formation. We estimate the collision-driven mass loss by combining the fitting formulas of \citet{glebbeek_structure_2013} with different analytic prescriptions and stellar-evolution calculations for the internal stellar structure.

\subsection{Simulation data}\label{sec:simulation_data}

The simulations of \citet{PSolar2025} were performed with the Astrophysical MUlti-purpose Software Environment \citep[AMUSE\footnote{https://github.com/amusecode/amuse}; see][]{PortegiesZwart2009,Pelupessy2013,PortegiesZwart2013,PortegiesZwart2018} framework. We modeled an embedded primordial protostellar cluster in the core of a minihalo. The simulations were carried out with the SPH code \textsc{Fi} \citep{Hernquist1989,Gerritsen1997} and the stellar dynamics with the pure N-body code \textsc{Ph4} \citep{McMillan1996}. To couple these two codes, we used the BRIDGE method \citep{Fujii2007}. The protostars were treated as sink particles and modeled according to \citet{hubber_improved_2013}. \citet{PSolar2025} systematically explored the impact of the initial gas temperature on the evolution of a protostellar cluster in a primordial gas cloud with a mass of $3\times10^{4}~\mathrm{M_{\odot}}$ and a virial radius of $0.14~\mathrm{pc}$, considering initial temperatures of $500,~1000,~3000,~5000,~$ and $8000~\mathrm{K}$. The simulations were designed assuming the direct-collapse scenario from \citet{latif_assessing_2015}, in which the gas evolves approximately isothermally under the influence of strong LW radiation that suppresses efficient \textrm{H2} cooling. To model the gas thermodynamics, they adopted a simplified equation of state (EoS), assuming an initially isothermal evolution that transitions to an adiabatic regime once the gas density exceeds $10^{15}~\mathrm{cm^{-3}}$.

The mass-radius relation of the protostars evolves according to the accretion rate, following relations from \citet{hosokawa_low-metallicity_2009,hosokawa_rapidly_2012,hosokawa_formation_2013}, who performed detailed stellar evolution calculations for a wide range of accreting protostars. These are one-dimensional stellar-evolution calculations that self-consistently solve the equations of stellar structure for a spherically symmetric, accreting protostar (see \autoref{sec:mass_loss_app}). We adopt this mass–radius relation because, unlike simplified analytic approximations, it remains physically motivated across the full range of accretion rates relevant to our simulations, from slowly accreting low-mass stars to rapidly accreting (super)massive protostars. Depending on the accretion rate, protostars evolve along the SMS ($\dot{m}\geq 0.04~\rm M_{\odot}~yr^{-1}$), VMS ($ 10^{-6}~\mathrm{M_{\odot}~yr^{-1}}\leq \dot{m}<0.04~\rm M_{\odot}~yr^{-1}$) or star ($\dot{m}< 10^{-6}~\rm M_{\odot}~yr^{-1}$) tracks. For further details of these evolutionary regimes and the adopted mass-radius relations, see Appendix A of \citet{reinoso_formation_2023}.

We found that the mass of the most massive object (MMO) increases when the gas temperature decreases, due to the increase in the initial instability of the gas cloud. The final masses of the MMOs are $\sim 19~000 - 28~000 ~\mathrm{M_{\odot}}$, from warmer to colder temperatures, after $\sim 30~000~\mathrm{yr}$. These values correspond to a final formation efficiency defined as the ratio between the final MMO mass and the total mass,
\begin{equation}\label{eq:efficiency}
    \varepsilon = M_{\rm MMO}/M_{\rm total},
\end{equation}
which exceeds $0.6$. General properties of the simulations performed in \citet{PSolar2025} are presented in \autoref{table:initial_data}.

 \begin{table}
 \centering
 %\small
 \caption{Summary of the data provided by \citet{PSolar2025}.}\label{table:initial_data}
 \begin{tabular}{lllll} 
 \hline 
 \hline
 Simulation &Temperature [K] & $M_{\rm MMO}~[\rm M_{\odot}]$ & $N_{\rm coll}$ & $\epsilon$ \\
 \hline
 $1$   & $8000$ &  $18766.54$ & $1051$ &  $0.63$ \\
 $2$   &$8000$ &  $18109.64$ & $1036$ & $0.60$\\
 $3$   &$5000$ & $21567.15$ & $1875$ &  $0.72$\\
 $4$   &$5000$ &  $21592.66$ & $1788$ &  $0.72$\\
 $5$   &$3000$ &  $23216.09$ & $2391$ &  $0.77$ \\
 $6$   &$3000$ &   $24993.19$ & $2710$ & $0.83$\\
 $7$   &$1000$ & $25563.54$ &$1770$ &  $0.85$\\
 $8$   &$1000$ &  $28789.49$ & $2720$ &  $0.96$\\
 $9$   &$500$ &  $28471.44$ & $1675$ &  $0.95$\\
 $10$   &$500$ &  $28875.27$ & $2880$ &  $0.96$\\
 \hline
 \hline
 \end{tabular}
 \tablefoot{Summary of the data provided by \citet{PSolar2025} showing initial gas temperature, final mass of the most massive object (MMO), number of collisions with the MMO and final efficiency of the MMO \pcref{eq:efficiency}.}
 \end{table}

 \begin{table}
 \centering
 %\small
 \caption{Summary of the enclosed-mass radii combinations and internal-structure prescriptions explored in this work.} \label{table:dif_stellar_radii}
 \begin{tabular}{lllll} 
 \hline 
 \hline
 Model & $86\%$ &$50\%$ & Internal-structure prescription \\
 \hline
 M1 & - & - & Lane-Emden (n=3)\\
 M2 & - & - & \citet{schleicher_2013}\\
 $\rm M3^1_{80-40}$ &80&40 &1-D data +Lane-Emden (n=3)\\
 $\rm M3^1_{80-50}$ &80&50 &1-D data +Lane-Emden (n=3)\\
 $\rm M3^1_{80-60}$ &80&60 &1-D data +Lane-Emden (n=3)\\
 $\rm M3^1_{82-50}$ &82&50 &1-D data +Lane-Emden (n=3)\\
 $\rm M3^1_{84-50}$ &84&50 &1-D data +Lane-Emden (n=3)\\
 $\rm M3^1_{86-50}$ &86&50 &1-D data +Lane-Emden (n=3)\\
 $\rm M3^2_{80-40}$ & 80 & 40 &1-D data +\citet{schleicher_2013}\\
 $\rm M3^2_{80-50}$ & 80 & 50 &1-D data +\citet{schleicher_2013}\\
 $\rm M3^2_{80-60}$ & 80 & 60 &1-D data +\citet{schleicher_2013}\\
 $\rm M3^2_{82-50}$ & 82 & 50 &1-D data +\citet{schleicher_2013}\\
 $\rm M3^2_{84-50}$ & 84 & 50 &1-D data +\citet{schleicher_2013}\\
 $\rm M3^2_{86-50}$ & 86 & 50 &1-D data +\citet{schleicher_2013}\\
 \hline
 \hline
 \end{tabular}
 \tablefoot{Column 2: enclosed-mass radius used to fit $86\%$ of the stellar mass; Column 3: enclosed-mass radius used to fit $50\%$ of the stellar mass; Column 4: internal-structure prescriptions used in the post-processing calculation. We combine the fitting formulas of \citet{glebbeek_structure_2013} with either the analytic prescriptions M1 and M2, or with the M3 prescription based on stellar-structure calculations of \citet{hosokawa_evolution_2009,hosokawa_evolution_2010,hosokawa_formation_2013} (see \autoref{sec:mass_loss_app}). To obtain the enclosed-mass radii across all stellar masses, and given the limited coverage of the one-dimensional stellar structure simulations, we use M1 or M2 to estimate the internal stellar structure outside the mass range covered by M3. Models based on stellar-evolution data are labeled according to their enclosed-mass radius combination.}
 \end{table}

\subsection{Mass-loss prescription}\label{sec:mass_loss_app}

In the context of SMBH formation, many studies assume mass conservation during stellar collisions, which leads to an overestimation of the final mass of the MMO. For rapidly accreting protostars, this makes collision-induced mass loss a key uncertainty in SMBH formation. To estimate the mass-loss fraction and the lower limit on MMO growth due to stellar collisions, we adopt the analytic mass-loss prescription used by \citet{glebbeek_structure_2013}. We apply their fitted formulas to our collision histories and combine them with different prescriptions to calculate the internal stellar structure. \citet{glebbeek_structure_2013} studied the long-term evolution of massive, head-on stellar mergers in the context of blue-straggler formation. Their stellar collisions involved massive main-sequence stars with primary masses in the range $5-40~\mathrm{M_{\odot}}$ and lower-mass companions. The collisions were simulated with \textsc{GADGET2} \citep{gadget2_2005} at a resolution of $262~000$ SPH particles. The merger products were evolved with the adaptive non-Lagrangian, non-Eulerian grid stellar-evolution code \textsc{STARS} \citep[e.g.][]{eggleton_1971,pols_1995,stancliffe_2006,glebbeck_2008}. The mass-loss approximation for low mass ratios ($q<0.4$) follows the model proposed by \citet{lombardi_stellar_2002}, who studied the collision of non-rotating low-mass stars. The fraction of mass lost in a collision depends on the internal structure of the parent stars according to the following equation:

\begin{equation}\label{eq:mass_loss_lombardi}
    \phi = C_1 \frac{q}{(1+q)^2}\frac{R_{1,0.86} +R_{2,0.86} }{R_{1,0.5} +R_{2,0.5}}~.
\end{equation}

Here, $q=M_2/M_1$ is the mass ratio (with $M_1>M_2$) and $R_{n,0.86}$ and $R_{n,0.50}$ are the radii enclosing $86\%$ and $50\%$ of the total mass of the parent star $n$. The parameter $C_1$ is a dimensionless constant with a value of $C_1=0.157$, calibrated from SPH calculations. For collisions with $q \geq 0.4$, the mass-loss fraction follows the fitting formula from \citet{glebbeek_2008_pols} as

\begin{equation}\label{eq:mass_loss_glebbeek}
    \phi = C_2 \frac{q}{(1+q)^2}~,
\end{equation}
with $C_2=0.3$. The new mass ($M_{\rm new}$) and the mass lost ($M_{\rm lost}$) of the merger are calculated as follows:

\begin{equation}
    M_{\rm new} = (M_1 + M_2)(1-\phi)~, \quad M_{\rm lost} = (M_1 + M_2) - M_{\rm new}.
\end{equation}

To evaluate the mass-loss prescriptions above, we require stellar radii at specific enclosed-mass coordinates (e.g. $R_{n,0.86}$ and $R_{n,0.50}$). We therefore consider three prescriptions for the internal stellar structure, based on analytic and numerical calculations. First, following the approach adopted in the mass-loss prescription of \citet{lombardi_stellar_2002}, we assume that the stars are in hydrostatic equilibrium and compute the internal structure of the colliding stars using a polytropic MS model described by the Lane–Emden equation \citep[e.g.][]{1870Lane,emden1907gaskugeln,1939Chandrasekhar}, which is obtained by adopting a polytropic EoS. In particular, $n = 3$ corresponds to a radiative MS star, while the limit $n \rightarrow \infty$ recovers an isothermal EoS. Hereafter, we refer to this prescription as M1. In the top panel of \autoref{fig:den_86_50_appendix}, we show the mass-radius relations at 86\% and 50\% enclosed mass for accreting protostars computed as MS stars. We adopt the mass-radius parametrization from \citet{reinoso_formation_2023}. 

We also model the stellar structure using the analytic accreting-protostar prescription of \citet{schleicher_2013}, who followed the evolution of contracting mass shells in rapidly accreting protostars, finding that for sufficiently high accretion rates ($\dot{m}\gg 0.14~\rm M_{\odot}~yr^{-1}$), the central object may collapse into a black hole while still embedded within an accreting envelope, leading to the formation of a quasi-star. At lower accretion rates ($\dot{m}\lesssim 0.14~\rm M_{\odot}yr^{-1}$), the object instead contracts gravitationally and evolves toward a main-sequence SMS. Following the approach adopted by \citet{schleicher_2013} and \citet{alister_seguel_formation_2020}, we obtain a mass–radius relationship for accreting primordial protostars:

\begin{equation}
    \label{eq:schleicher_2013}
    \frac{1000 ~\mathrm{R_{\odot}}}{R} = \frac{1000}{260(M/\mathrm{M_{\odot}})^{1/2}} + \frac{1.04~\mathrm{yr^{-1}}[t-t_{\mathrm{ini}}(M)]}{M/\mathrm{M_{\odot}}}~.
\end{equation}

Each term on the right-hand side of \cref{eq:schleicher_2013} depends on the characteristic timescale of the protostar. To determine the radius enclosing a mass fraction $i$ of the parent star $n$, we consider two limiting cases depending on whether the accretion timescale ($t_{\rm acc}$) or the Kelvin-Helmholtz timescale ($t_{\rm KH}$) dominates the stellar evolution. At low accretion rates ($\dot{m}\lesssim 0.14~\rm M_{\odot}yr^{-1}$), protostars contract more efficiently because the transition from the $t_{\rm acc}$-dominated regime to the $t_{\rm KH}$-dominated regime occurs earlier at lower stellar masses. When the accretion timescale dominates the protostellar evolution ($t_{\rm acc}\ll t_{\rm KH}$), the first term on the right-hand side of \cref{eq:schleicher_2013} dominates after the formation of a mass shell and therefore determines its initial radius \citep[][]{schleicher_2013,alister_seguel_formation_2020}. In this regime, the characteristic radius $R_i$ is given by

\begin{equation}
    R_{n,i}=260 \left( \frac{i \cdot M_n }{\mathrm{M_{\odot}}} \right)^{1/2} \mathrm{R_{\odot}}~.
    \label{eq:tkh_ll_tacc}
\end{equation}

After this initial growth phase, the radius and luminosity increase by several orders of magnitude, leading to an inversion between $t_{\rm acc}$ and $t_{\rm KH}$. Therefore, the second term of \cref{eq:schleicher_2013} dominates the stellar evolution. When Kelvin-Helmholtz contraction dominates the evolution ($t_{\rm KH}\ll t_{\rm acc}$), we compute the stellar structure using

 \begin{equation}
    R_{n,i}=\frac{i \cdot M_n }{1.04 \times 10^{-3}~\mathrm{M_{\odot}}} \frac{1}{t_{\rm KH}/\mathrm{yr}} ~ \mathrm{R_{\odot}}~.
    \label{eq:tacc_ll_takh}
\end{equation}

In the bottom panel of \autoref{fig:den_86_50_appendix}, we present the mass-radius relation for this prescription, showing the radii enclosing $86\%$ and $50\%$ of the stellar mass for different accretion rates, as described by \cref{eq:tkh_ll_tacc} and \cref{eq:tacc_ll_takh}. Hereafter, we refer to this prescription as M2.

\begin{figure}[H]
    \begin{subfigure}[b]{\linewidth}
        \centering
        \panellabel[9mm]{0.9\linewidth}{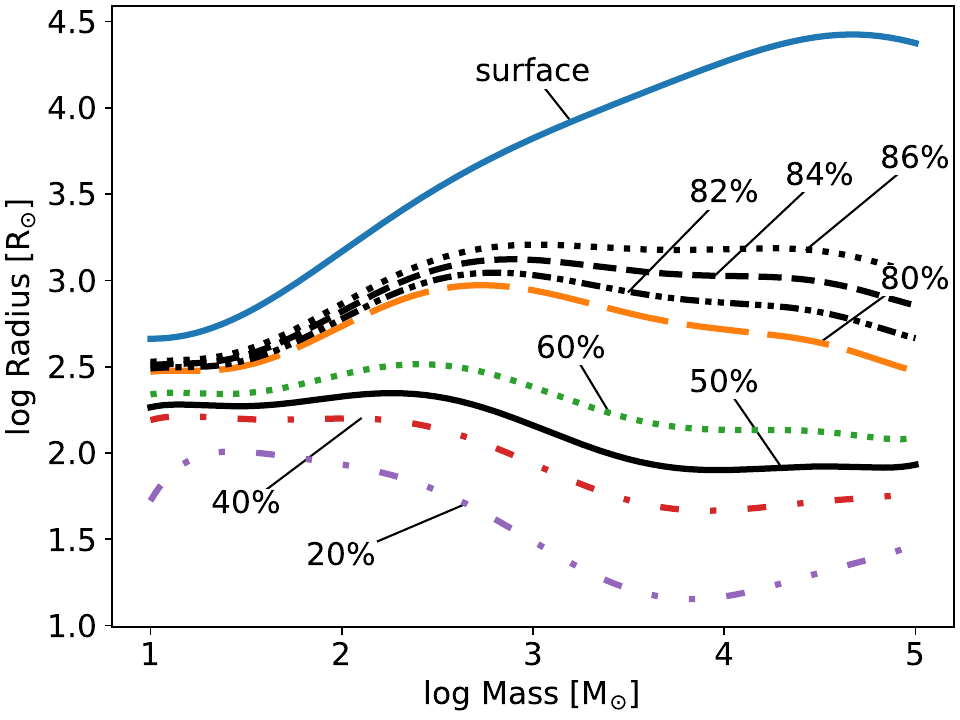}{(a)}
        \phantomsubcaption\label{fig:SMS structure}
    \end{subfigure}

    \begin{subfigure}[b]{\linewidth}
        \centering
        \panellabel[9mm]{0.9\linewidth}{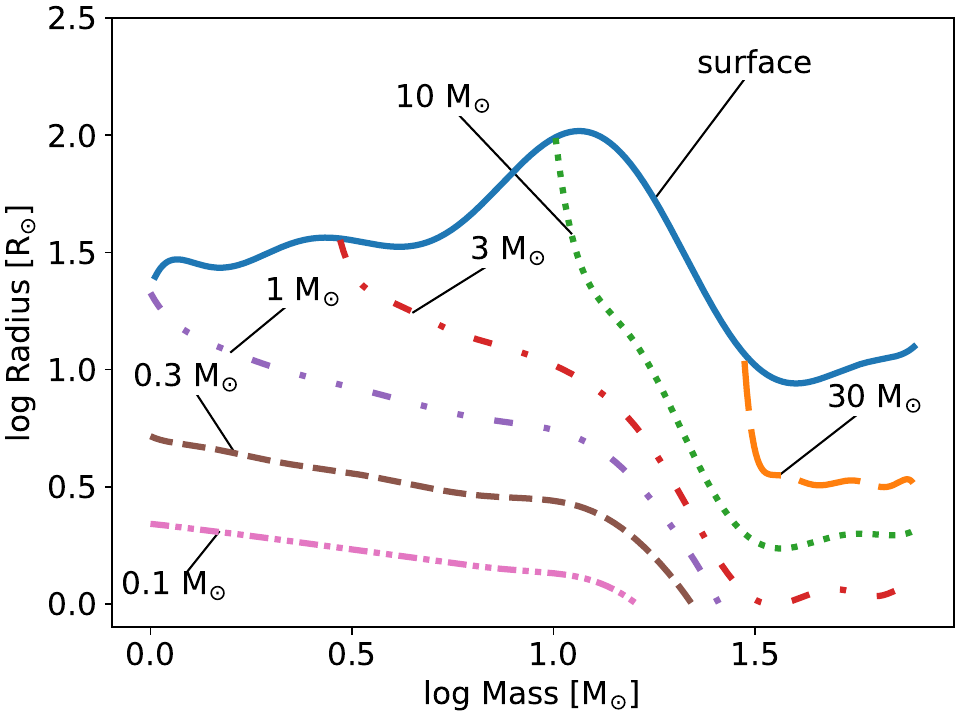}{(b)}
        \phantomsubcaption\label{fig:VMS structure}
    \end{subfigure}

    \begin{subfigure}[b]{\linewidth}
        \centering
        \panellabel[13.mm]{0.9\linewidth}{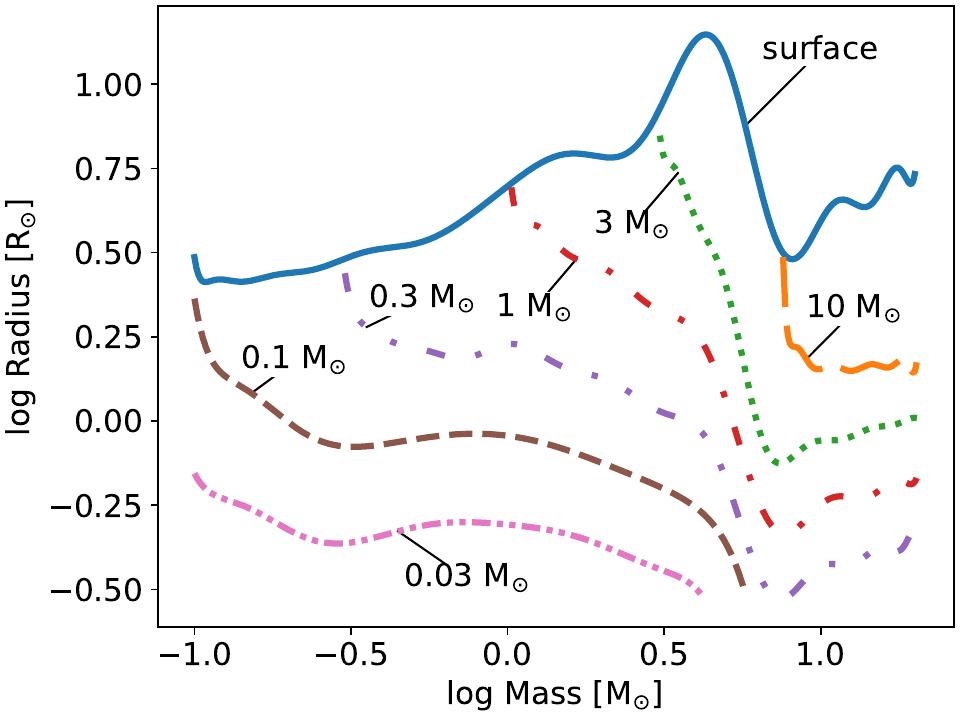}{(c)}
        \phantomsubcaption\label{fig:NORMAL structure}
    \end{subfigure}
    \caption{Mass-radius evolution to calculate internal structure of protostars along SMS, VMS and star tracks in the collision history. Panel a: Mass-radius relation based on \citet{hosokawa_formation_2013} for a protostar with an accretion of $\dot{m} = 1.0~\mathrm{M_{\odot}~yr^{-1}}$ over the mass range $M\sim 10-10^5~\rm M_{\odot}$. The purple, red, green and orange curves show the radii enclosing $20\%$, $40\%$, $60\%$ and $80\%$ of the stellar mass, respectively. The black curves indicate interpolated radii enclosing $50\%$, $82\%$, $84\%$ and $86\%$ of the stellar mass. Panel b: Mass-radius evolution based on \citet{hosokawa_evolution_2010} for a protostar with an accretion of $\dot{m} = 10^{-3}~\mathrm{M_{\odot}~yr^{-1}}$ over the mass range $M\sim 1-10^{1.75}~\rm M_{\odot}$. The pink, brown, purple, red, green and orange curves show the locations of fixed mass ($M=0.1$, $0.3$, $1$, $3$, $10$ and $30~\mathrm{M_{\odot}}$). Panel c: Mass-radius evolution based on \citet{hosokawa_evolution_2009} for a protostar with an accretion of $\dot{m} = 10^{-5}~\mathrm{M_{\odot}~yr^{-1}}$ over the mass range $M\sim 0.1-10^{1.3}~\rm M_{\odot}$. The pink, brown, purple, red, green and orange curves show the locations of fixed mass ($M=0.03$, $0.1$, $0.3$, $1$, $3$ and $10~\mathrm{M_{\odot}}$). For all panels, the solid blue lines show the stellar surface radius.}
\end{figure}

Finally, we compute the internal structure using one-dimensional (1-D) stellar-evolution simulations for accreting protostars from moderate to high accretion rates $10^{-6}~\mathrm{M_{\odot}~yr^{-1}} \leq \dot{m}\leq 0.04~\mathrm{M_{\odot}~yr^{-1}}$ \citep[e.g.][]{hosokawa_evolution_2009,hosokawa_evolution_2010,hosokawa_formation_2013}. Unlike full hydrodynamical simulations, 1-D stellar evolution simulations solve the four stellar structure equations, i.e., the equations of continuity, hydrostatic equilibrium, energy conservation, and energy transfer, assuming spherical symmetry. Under these conditions, the internal structure of accreting protostars may differ substantially from that of MS stars, making it necessary to resolve the stellar structure over a broad range of accretion rates. We use these data to calculate the internal structure for the three main evolutionary tracks (SMS, VMS, and star) adopted in \citet{PSolar2025}, differentiated by their characteristic accretion rates. These include an inflated SMS track for $\dot{m}\geq 0.04~\rm M_{\odot}~yr^{-1}$ (\autoref{sec:SMS_track}), a VMS track for  $10^{-6}~\rm M_{\odot}~yr^{-1}\leq \dot{m}<0.04~\rm M_{\odot}~yr^{-1}$ (\autoref{sec:VMS_track}), and a normal stellar track for $\dot{m}< 10^{-6}~\rm M_{\odot}~yr^{-1}$(\autoref{sec:normal_track}). For each evolutionary track, we use the published mass-radius evolution at the stellar surface and at internal mass coordinates to estimate the radii enclosing 50\% and 86\% of the stellar mass, which are required by our mass-loss prescription. As the evolutionary track of a colliding star depends on its accretion rate, it may transition between different structural regimes over its lifetime, which could change its mass-loss history. Hereafter, we refer to this prescription as M3.

In the $\rm M3$ prescription, we cannot follow the full evolution of every collider because the available stellar-structure data cover only a limited range of masses on each accretion track. We therefore adopt model M1 or model M2 to compute the internal structure for colliders outside the simulated mass range. For example, for stars with $\dot{m}<0.04~\rm M_{\odot}~yr^{-1}$ and $M \gtrsim 56~\rm M_{\odot}$, we approximate their internal structure using either the M1 or M2 prescription. For clarity, we denote prescriptions based on 1-D stellar-structure simulations as $\rm M3^A_{B-C}$, where $A$ denotes the theoretical model used to compute the stellar structure outside the range of validity of the 1-D  stellar-structure data, while $B$ and $C$ indicate the adopted enclosed-mass radii (e.g. $\rm M3^1_{86-50}$). All explored models are summarized in \autoref{table:dif_stellar_radii}.

\subsubsection{SMS track}\label{sec:SMS_track}

Protostars with $\dot{m}\geq 0.04~\rm M_{\odot}~yr^{-1}$ are on the SMS track and follow a mass-radius relation given by

\begin{equation}\label{eq:sms}
    R_{\rm star} = 2600\left( \frac{M_{\rm star}}{100~\rm M_{\odot}} \right)^{1/2}~\rm R_{\odot}~.
\end{equation}

We compute the internal structure along this track using the rapidly accreting stellar models of \citet{hosokawa_formation_2013}, with an accretion rate of $\dot{m}= 1.0~\mathrm{M_{\odot}~yr^{-1}}$ and a fraction of the accretion luminosity deposited in the stellar interior of $\eta = 0.1$. With the stellar evolution code from \citet{Yorke_2008}, they followed the evolution of the mass-radius relation for the stellar surface and mass coordinates corresponding to 80\%, 60\%, 40\%, and 20\% of the total stellar mass. They showed that a star becomes supermassive while it is in the supergiant protostar stage, during which the star exhibits a bloated envelope and a contracting inner core. In \autoref{fig:SMS structure}, we present the fit used to estimate the internal structure of the colliding stars along the SMS track, following the results of \citet{hosokawa_formation_2013}. Because these mass coordinates do not exactly match those required in \cref{eq:mass_loss_lombardi}, we use logarithmic interpolation to obtain $R_{0.5}$ and $R_{0.82}$.

\subsubsection{VMS track}\label{sec:VMS_track}

Protostars on the VMS track have accretion rates in the range $10^{-6}-0.04~\mathrm{M_{\odot}~yr^{-1}}$ and follow an evolution characterized by three phases: the adiabatic accretion phase, the swelling phase, and the Kelvin-Helmholtz contraction phase \citep{hosokawa_low-metallicity_2009}. For this track, we compute the internal structure using the results of \citet{hosokawa_evolution_2010}, who modeled the internal evolution of spherically accreting protostars at $\dot{m}= 10^{-3}~\mathrm{M_{\odot}~yr^{-1}}$ using the calculation method developed by \citet{Stahler_1986} and \citet{PallaStahler1991}.

In \autoref{fig:VMS structure}, we show the internal structure used to calculate the mass-loss fraction. Unlike the SMS case, the curves correspond to fixed mass coordinates of $M=0.1$, $0.3$, $1$, $3$, $10$ and $30~\mathrm{M_{\odot}}$. We therefore cannot directly obtain the radii enclosing 86\% and 50\% of the total stellar mass. Instead, for a given protostellar mass, we adopt the radii at the mass coordinates closest to 86\% and 50\% of the enclosed mass.

\subsubsection{Normal track}\label{sec:normal_track}

Protostars with low accretion rates ($\dot{m}\leq 10^{-6}~\mathrm{M_{\odot}~yr^{-1}}$) follow the normal evolutionary track, according to the models used by \citet{reinoso_formation_2023} and \citet{PSolar2025}. We estimate the internal structure using the data presented by \citet{hosokawa_evolution_2009}. In particular, we use the evolution of a protostar accreting at a constant rate of $\dot{m} = 10^{-5}~\mathrm{M_{\odot}~yr^{-1}}$. The corresponding stellar structure is shown in \autoref{fig:NORMAL structure}, where the curves denote the evolution of fixed mass coordinates of $M=0.03$, $0.1$, $0.3$, $1$, $3$ and $10~\mathrm{M_{\odot}}$. As for the VMS track, we use the same procedure to determine the radii enclosing 86\% and 50\% of the stellar mass.

\subsection{Numerical methods} \label{subsect:numerical_methods}

In this work, we compute the mass lost in each stellar collision using the masses, radii, and an approximation of the internal stellar structure at the time of the merger. We trace the full collisional history to obtain the cumulative mass loss and the final mass of the CMO for different gas temperatures. Due to the limitations of the post-processing approach, we cannot estimate the impact of the mass reduction on the gas accretion and cluster dynamics, and therefore we assume the same accretion history for all mergers. Since the data on the internal structure of rapidly accreting stars are not available in electronic form, we digitized them using the WebPlotDigitizer tool\footnote{https://automeris.io} \citep{WebPlotDigitiser_paper}. Polynomial terms used to build the stellar profiles are available at \href{https://github.com/pasolar/Collision-driven-mass-loss.git}{this repository}. Before applying this method to the full simulations in \autoref{sec:results}, we first examine isolated collisions under simplified, constant accretion rates in \autoref{sec:analytical_approach}, to characterize how the prescriptions compare in a controlled setting.

\begin{figure*}
    \begin{subfigure}[b]{\linewidth}
        \centering
        \includegraphics[width=0.9\linewidth]{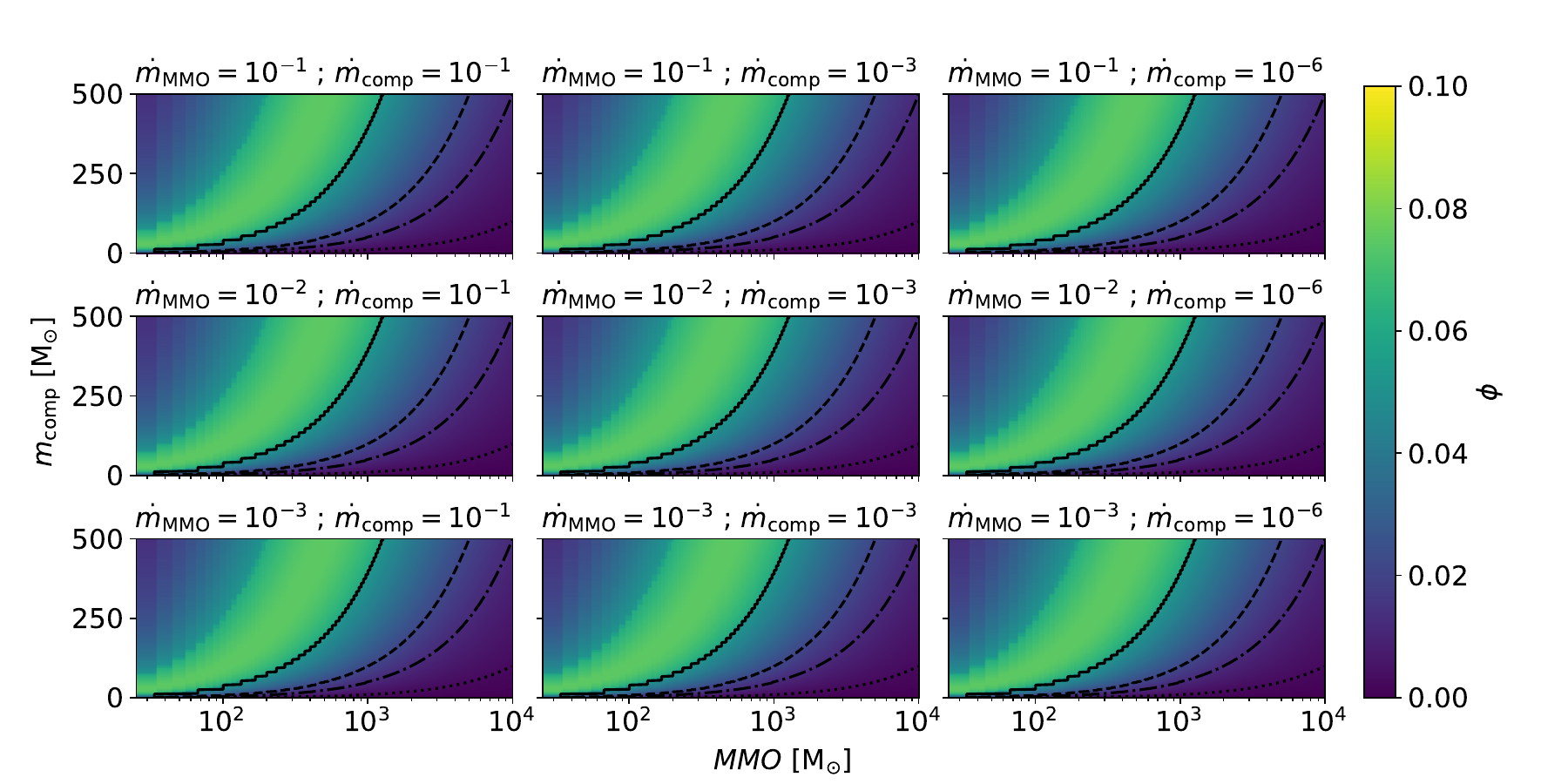}
    \caption{}%\textbf{Mass-loss fraction computed using  the analytic Lane-Emden stellar structure prescription (model M1).}}
    \label{fig:acc_den_mass_loss_poly}
    \end{subfigure}
    \hfill
    \begin{subfigure}[b]{\linewidth}
        \centering
        \includegraphics[width=0.9\linewidth]{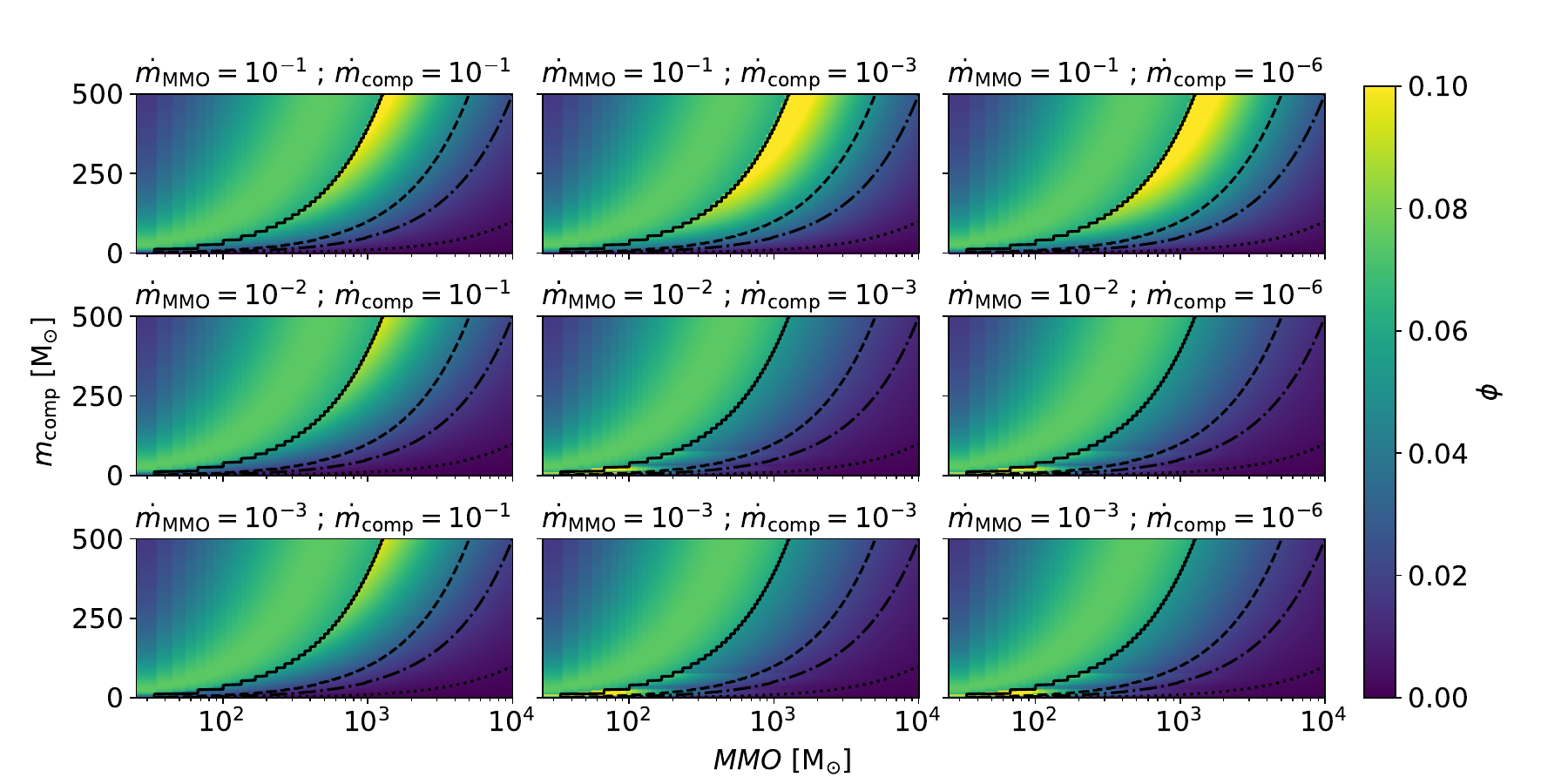}
    \caption{}%\textbf{Mass-loss fraction computed using stellar structure derived from 1-D stellar evolution calculations ($\rm M3^1_{80-60}$).}}
    \label{fig:acc_den_mass_loss_stellar}
    \end{subfigure}
    \caption{Color map of the mass-loss fraction as a function of the accretion rates of the most massive object (MMO) and the companion star. Panel (a) shows results obtained with the analytic stellar structure prescription M1, based on Lane-Emden polytropic models, while panel (b) uses the $\rm M3^1_{80-60}$ prescription based on 1-D stellar evolution calculations of accreting protostars. Each row corresponds to a different accretion rate of the MMO, $\dot{m}_{\rm MMO} = 10^{-1},10^{-2}~\mathrm{and}~10^{-3}~\rm M_{\odot}~yr^{-1}$, while each column corresponds to companion accretion rates of $\dot{m}_{\rm comp}=10^{-1},10^{-3}~\mathrm{and}~10^{-6} ~\rm M_{\odot}~yr^{-1}$. The color scale indicates the mass-loss fraction in each isolated collision. The solid, dashed, dash-dotted, and dotted black curves correspond to mass ratios of $q=0.4$, $0.1$, $0.05$, and $0.01$. Although both prescriptions predict similar global trends, the model $\rm M3^1_{80-60}$ yields systematically larger mass loss in the high-accretion SMS regime because rapidly accreting stars develop more extended and weakly bound envelopes than the analytic model M1.}
    \label{fig:color_map_teo_all}
\end{figure*}

\section{Collision-driven mass loss - Analytical approach}\label{sec:analytical_approach}

In SMBH formation scenarios, stellar collisions may occur between parent stars experiencing different episodic accretion rates and therefore in distinct evolutionary stages. To estimate the collision-driven mass loss across different accretion and mass regimes, we compare the predictions obtained with models M1, M2, and $\rm M3^1_{80-60}$ for isolated stellar collisions between an SMS and a companion star accreting at different rates. In this analysis, we do not consider time-dependent or variable accretion rates. Instead, we model a set of constant accretion rates representative of episodic accretion events expected in SMBH formation pathways. In scenarios involving gas accretion and runaway stellar collisions, protostars may accrete over a broad range of rates, from $\sim 10^{-6}~\rm M_{\odot}~yr^{-1}$ up to $\sim1~\rm M_{\odot}~yr^{-1} $ \citep[][]{Chon_2020,chon_impact_2022,reinoso_formation_2023,nandal_critical_2023,PSolar2025}. This naturally leads to collisions between stars accreting at substantially different accretion rates.

Comparing the analytic models M1 and M2, we find only small differences in the collision-driven mass loss along the MMO evolution for all combinations of accretion rates. M2 yields larger mass loss at any phase of the colliding stars because it predicts a less compact stellar structure, leading to up to $\sim 4.9\% $ more mass loss than M1. However, when the massive protostar evolves along the VMS track and the companion lies on the SMS track, M1 predicts up to $\sim18\%$ more mass loss than M2. Given the relatively small differences between the two analytic prescriptions, we adopt M1 as a representative case for models based on analytically motivated stellar structures. Nevertheless, depending on the characteristic stellar masses and accretion rates during the runaway-collision phase, these differences may still affect the final mass of the MMO.

In \autoref{fig:color_map_teo_all}, we present color maps of the mass-loss fraction for stellar collisions involving different accretion rates, comparing the predictions of models M1 and $\rm M3^1_{80-60}$ (a similar figure for model M2 is presented in \autoref{appendix:extra_figures}). We find significant differences between the two models: M1 predicts the same mass-loss fraction across all accretion rates, whereas $\rm M3^1_{80-60}$ predicts a higher mass-loss fraction when the MMO lies on the SMS track, where the maximum mass-loss fraction for M1 is $7.5\%$ and $27.8\%$ for $\rm M3^1_{80-60}$. Compared to the analytic prescriptions M1 and M2, the less compact structure of model $\rm M3^1_{80-60}$ leads to larger collision-driven mass loss at smaller mass ratios $q$.

 \begin{table*}
 \centering
 \small
 %\footnotesize
 \setlength{\tabcolsep}{3pt}
 \caption{Results obtained using different internal-structure prescriptions.} \label{table:results_model_1}
 %\begin{tabular}{lllllllllllllllllll} 
 \begin{tabular}{l|ccccc|ccccc|ccccc} 
 \hline 
 \hline
 & \multicolumn{5}{c|}{\subcaptionbox{Model M1\label{table:results_model_1}}[.1\textwidth]{}}
 & \multicolumn{5}{c|}{\subcaptionbox{Model M2\label{table:results_model_2}}[.1\textwidth]{}}
 & \multicolumn{5}{c}{\subcaptionbox{Model $\rm M3^1_{80-60}$\label{table:results_model_3}}[.2\textwidth]{}}\\
 %\cline{2-16}
 Sim&$M_\mathrm{{MMO,post}}$ & $M_\mathrm{{lost,tot}}$& $M_\mathrm{{lost,MMO}}$ &\% lost & $\epsilon_{\mathrm{post}}$& $M_\mathrm{{MMO,post}}$ & $M_\mathrm{{lost,tot}}$& $M\mathrm{_{lost,MMO}}$ &\% lost & $\epsilon_{\mathrm{post}}$& $M\mathrm{_{MMO,post}}$ & $M_\mathrm{{lost,tot}}$& $M_\mathrm{{lost,MMO}}$ &\% lost & $\epsilon_{\mathrm{post}}$
 \\
  %&$\times10^3 \mathrm{[M_{\odot}]}$& $\times10^3\mathrm{[M_{\odot}]}$&$\times10^3\mathrm{[M_{\odot}]}$ & &&$\times10^3\mathrm{[M_{\odot}]}$& $\times10^3\mathrm{[M_{\odot}]}$&$\times10^3\mathrm{[M_{\odot}]}$ & &&$\times10^3\mathrm{[M_{\odot}]}$& $\times10^3\mathrm{[M_{\odot}]}$&$\times10^3\mathrm{[M_{\odot}]}$& &  \\
 \hline
 $1$   & $14.09$ & $4.72$& $4.67$&$24.91$ &$0.47$& 13.90 & 4.92 & 4.87 &25.96 &0.46& 11.33 & 7.49 & 7.44 & 39.64 &0.38  \\
 $2$   & $14.37$ & $3.76$ &$3.74$&$20.63$ &$0.48$& 14.21 & 3.92 & 3.90 &21.53 &0.47 & 12.11 & 6.03 & 6.00 &33.14 &0.40\\
 $3$   & $17.52$ & $4.06$ & $4.05$&$18.79$ &$0.58$& 17.33 & 4.25 & 4.24 &19.65 &0.58 & 14.44 & 7.13 & 7.12 &33.02 &0.48\\
 $4$   & $17.09$ & $4.53$ & $4.51$&$20.87$ &$0.57$& 16.88 & 4.74 & 4.71 &21.81 &0.56 & 13.93 & 7.69 & 7.66 &35.47 &0.46\\
 $5$   & $18.41$ & $4.91$& $4.81$& $20.71$&$0.61$& 18.19 & 5.13 & 5.03 &21.66 &0.61& 14.64 & 8.70 & 8.58& 36.95 &0.49  \\
 $6$   & $20.06$ & $5.01$ &$4.93$& $19.73$&$0.67$& 19.83 & 5.24 & 5.16 &20.64 &0.66 & 15.82 & 9.26 & 9.17 & 36.70&0.53\\
 $7$   & $23.24$ &$2.32$ & $2.32$&$9.07$ &$0.77$& 23.13 & 2.43 & 2.43 &9.50 &0.77 & 20.94 & 4.62 & 4.62   &18.09 &0.70\\
 $8$   & $25.15$ & $3.64$ & $3.64$&$12.63$ &$0.84$& 24.98 & 3.81 & 3.81 &13.24 &0.83 & 21.43 & 7.36 & 7.36 &25.55 &0.71\\
 $9$   & $26.58$ & $1.89$ & $1.89$&$6.65$ &$0.89$& 26.49 & 1.98 & 1.98 &6.97 &0.88 & 24.56 & 3.91 & 3.91 &13.72 &0.82\\
 $10$  & $25.32$ & $3.56$ & $3.56$&$12.33$ &$0.84$& 25.15 & 3.73 & 3.73 &12.92&0.84 & 21.57 & 7.31 & 7.31 &25.30 &0.72\\
 \hline
 \hline
 \end{tabular}
 \tablefoot{Summary of MMO results for different temperatures using different internal-structure prescriptions (models M1, M2 and $\rm M3^1_{80-60}$) to compute the collision-driven mass loss. We report the final post-mass-loss MMO mass ($M_{\rm MMO,~post}$), the total mass lost by stellar collisions ($M_{\rm lost,~tot}$), the total mass lost by the MMO ($M_{\rm lost,~MMO}$), the MMO mass-loss fraction ($\% \rm lost$) and the post-mass-loss formation efficiency ($\epsilon_{\rm post}$), computed as in \cref{eq:efficiency}. Masses of $M_{\rm MMO,~post}$, $M_{\rm lost,~tot}$ and $M_{\rm lost,~MMO}$ are in units of $10^3~\mathrm{M_{\odot}}$.}
\end{table*}

\section{Data analysis} \label{sec:results}

In this section, we use a post-processing analysis to compare the simulations of \citet{PSolar2025} using different mass-loss models. \citet{PSolar2025} systematically explored the impact of the initial gas temperature ($500 -8000~\mathrm{K}$) on the evolution of a protostellar cluster in a primordial gas cloud with a mass of $3\times10^{4}~\mathrm{M_{\odot}}$ and a virial radius of $0.14~\mathrm{pc}$ (see \autoref{sec:simulation_data}). We compute the mass loss of the MMO by (i) fitting the internal structure of the colliding stars as MS stars (M1), (ii) adopting the \citet{schleicher_2013} prescription based on accretion and Kelvin-Helmholtz timescales (M2), and (iii) modeling the internal structure from stellar-evolution simulations for rapidly accreting protostars using $80\%$ and $60\%$ of the enclosed-mass radii to compute the internal structure along the SMS track ($\rm M3^1_{80-60}$). For simplicity, in this section we use model M1 to compute stellar structures outside the mass range covered by the stellar-evolution data; a corresponding analysis using M2 is presented in \autoref{appendix:m3_2_model_cases}. We summarize the collision-driven mass loss for M1, M2, and $\rm M3^1_{80-60}$ in Tables 3a, 3b, and 3c.

\begin{figure}
    %\centering
    \includegraphics[width=\linewidth]{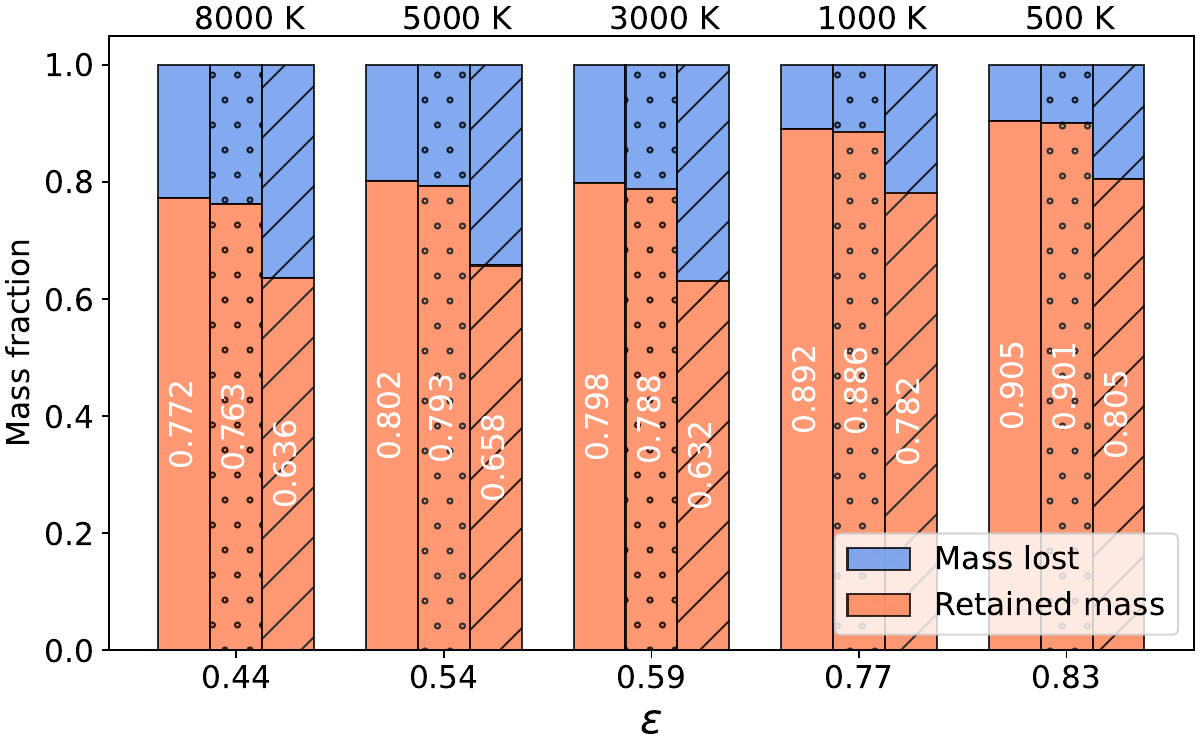}
    \caption{Average mass-loss fraction per collision and retained mass of the MMO for different initial gas temperatures as a function of the final MMO efficiency. Solid bars: mass loss with M1; dotted bars: mass loss with M2; crossed-out bars: mass loss with $\rm M3^1_{80-60}$.}
    \label{fig:mass_loss1}
\end{figure}

We find a clear trend in the mass-loss fraction for simulations with initial gas temperature $T\geq 3000~\mathrm{K}$, where the MMO loses $\sim 21\%$ of its final mass in M1, $\sim 22\%$ in M2, and $\sim 36\%$ in $\rm M3^1_{80-60}$. For simulations at $T< 3000~\mathrm{K}$, the mean mass loss is $\sim 10.17\%$ for M1, $\sim 10.7\%$ for M2 and $\sim 20.7\%$ for $\rm M3^1_{80-60}$. The two analytic prescriptions yield similar total mass loss. Relative to M1 and M2, $\rm M3^1_{80-60}$ predicts a mass-loss fraction that is $41.5 \pm 3.3\%$ and $38.9 \pm 3.3\%$ higher for simulations with $T\geq 3000~\mathrm{K}$; for $T< 3000~\mathrm{K}$, it yields $50.8 \pm 0.6 \%$ more mass loss than M1 and $48.4\pm 0.7 \%$ more than M2. In \autoref{fig:mass_loss1}, we show the average mass-loss fraction for each temperature and model as a function of the final MMO efficiency \pcref{eq:efficiency}. Colder simulations ($T< 3000~\mathrm{K}$) exhibit lower mass loss, reaching a final efficiency of $0.83 \pm0.056$ for simulations with $500~\rm K$, whereas simulations with $8000~\rm K$ reach a lower final efficiency of $0.44 \pm0.039$. This behavior may be related to the mass ratio, $q$, in \cref{eq:mass_loss_lombardi} and \cref{eq:mass_loss_glebbeek}, where the typical stellar mass of the stars colliding with the MMO is $\sim 15\rm~M_{\odot}$ at $8000~\rm K$ and $\sim 5\rm~M_{\odot}$ at $500~\rm K$. Thus, warmer simulations ($T\geq 3000~\mathrm{K}$) involve more massive colliders, which contribute more to MMO growth but also lead to larger mass-loss fractions.

We also need to consider that the main mechanism by which the MMO grows for simulations with $T\geq 3000~\mathrm{K}$ is stellar collisions, which contribute $75.16\pm 4.60\%$ to the final MMO mass. By contrast, in colder simulations, the predominant contribution is gas accretion, reaching $60.50\pm 9.55\%$ and thereby reducing the impact of collision-driven mass loss. From \autoref{fig:mass_loss1}, we find that $\rm M3^1_{80-60}$ predicts more mass loss per collision than the analytic models M1 and M2. This suggests that simplified structure models tend to underestimate collision-driven mass loss, even considering a relatively compact choice of the internal radii along the SMS track. Once the SMSs exceed $\sim 200~\rm M_{\odot}$, the Kelvin-Helmholtz timescale becomes shorter than the accretion timescale ($t_{\rm KH} < t_{\rm acc}$), triggering core contraction \citep[e.g.][]{hosokawa_rapidly_2012,hosokawa_formation_2013}. This structural change increases the ratio $(R_{1,0.86}+R_{2,0.86})/(R_{1,0.5}+R_{2,0.5})$ relative to the values obtained with the analytic prescriptions, resulting in an enhanced mass-loss fraction (up to a factor of $\sim2.7$ larger).

In addition to the mass ratio, $q$, we must consider the impact of the collision history on the final MMO mass. Although most stars experience their first collision directly with the MMO, some of them collide with each other before colliding with the MMO, losing mass in the process. Comparing the cumulative mass loss of the MMO with the mass loss from direct collisions alone, we find an average difference of $\sim 5.3 \pm 0.1\%$ for $T= 8000~\rm K$ and $\sim 0.05 \pm 0.001\%$ for $T=500~\rm K$ across all mass-loss models. This difference is attributed to mass lost in earlier collisions. In colder simulations, this effect becomes negligible, particularly in systems closer to a supercompetitive accretion scenario, although in less unstable systems additional prior collisions may reduce the mass available for the MMO. This effect may be more significant in rapidly accreting protostellar clusters, where we can expect more collisions that do not involve the MMO.

\begin{figure}
    %\centering
    \includegraphics[width=\linewidth]{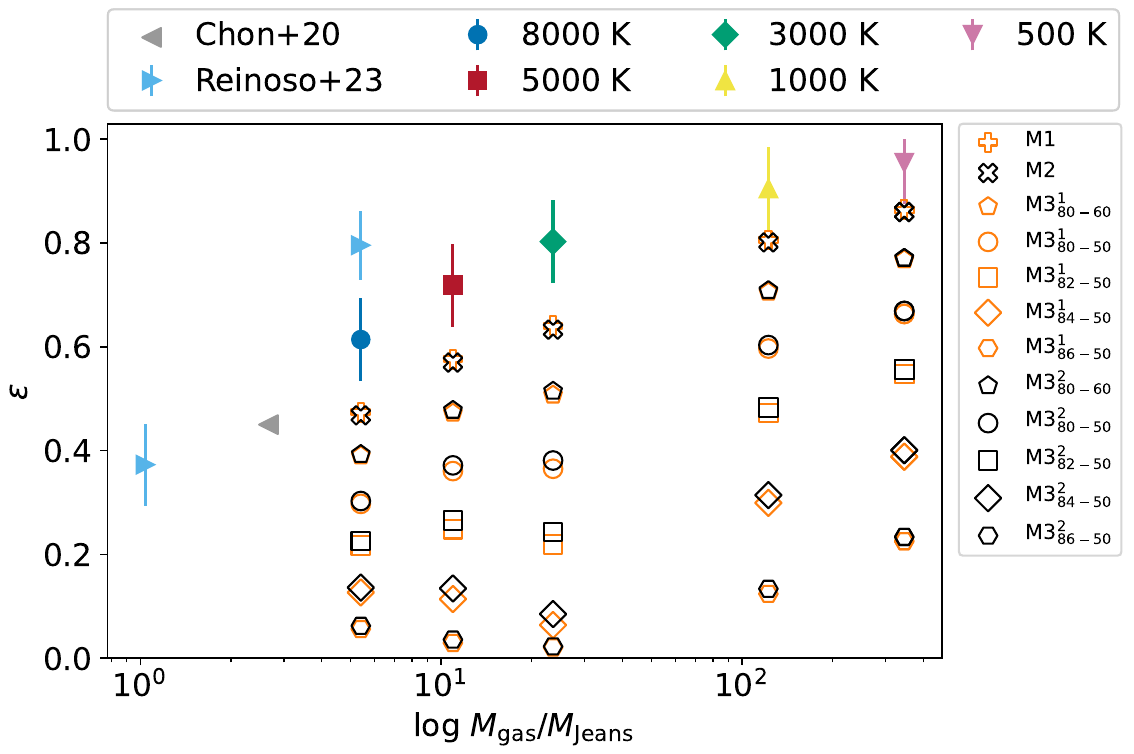}
    \caption{Efficiency (\textbf{$\varepsilon$}) of forming an MMO (defined in \cref{eq:efficiency}, as the ratio of mass in that object over total mass) considering collision-induced mass loss as a function of the gas mass divided by the thermal Jeans mass. We consider mass loss per collision by all cases listed in \cref{table:dif_stellar_radii}. We include data points from simulations that provide detailed models including collisions at sub-solar metallicities \citep{Chon_2020,reinoso_formation_2023,PSolar2025}.}
    \label{fig:efficience}
\end{figure}

Mass loss in stellar collisions affects the final efficiency, $\varepsilon$, of the MMO, as defined in \cref{eq:efficiency}. In \autoref{fig:efficience}, we present the efficiency, \textbf{$\varepsilon$}, of the MMO, as a function of the ratio between the gas mass ($M_{\rm gas}$) and the thermal Jeans mass ($M_{\rm Jeans}$). The collision-driven mass loss can notably decrease the MMO efficiency, reaching a maximum decrease of $\sim 25.4\%$, $\sim 26.9\%$ and $\sim 39.7\%$ for M1, M2, and $\rm M3^1_{80-60}$, respectively. As discussed above, the contribution by collisions to the final MMO mass decreases at lower temperatures; consequently, the reduction in mass due to collision-driven mass loss is smaller. For a gas cloud at $500~\rm K$, we find average efficiency reductions of $9.4\%$, $10.5\%$ and $19.9\%$ for M1, M2 and $\rm M3^1_{80-60}$, respectively. Thus, in supercompetitive accretion regimes or gas-dominated scenarios where accretion is the dominant growth channel, we expect that mass loss in collisions has only a minor impact. Across all simulations, $\rm M3^1_{80-60}$ yields the largest mass-loss effects because higher accretion rates imply less compact stellar structures, increasing the ratio $(R_{1,0.86}+R_{2,0.86})/(R_{1,0.5}+R_{2,0.5})$ over a broad range of $q$.

In addition, collision-driven mass loss may affect the overall cluster evolution on the Kelvin-Helmholtz timescale of the merger. After the collision, the shock heating increases the entropy of the merger product, causing it to expand, since it is initially out of thermal equilibrium, starting a bloated state. After this first expansion, as the core is not yet sufficiently hot or dense, the merger collapses on a Kelvin-Helmholtz timescale to reach thermal equilibrium again. This can lead to an increase of the collision probability of the MMO, as well as more weakly bound envelopes, increasing the probability of mass loss in subsequent close encounters (additional collisions or TDEs) \citep[e.g.][]{sills_evolution_1997,dale_collisions_2006,davies_stellar_2006,glebbeek_structure_2013}.

\section{Mass-loss fraction for different stellar structures} \label{subsect:stellar structure results}

\cref{eq:mass_loss_lombardi} requires the radii enclosing $86\%$ and $50\%$ of the stellar mass of the colliding stars to compute the mass-loss fraction. However, our dataset does not provide radii at these exact enclosed-mass coordinates; for the SMS track, we only have profiles at 80\%, 60\% and 40\% enclosed mass. In this section, we explore the sensitivity of the mass-loss fraction to different selections of the internal structure of the colliding stars, as well as the resulting impact on the final efficiency of the MMO. We estimate the radii enclosing 50\%, 82\%, 84\% and 86\% of the stellar mass via interpolation. Interpolating to 50\% enclosed mass should provide a reasonable approximation to the real internal structure. By contrast, estimating the radius at 82\%, 84\% and 86\% of the enclosed mass is more uncertain because of the complex evolution of the predominantly convective outer layers of collision remnants. The black solid, double-dot-dashed, dashed, and dotted curves in \autoref{fig:SMS structure} denote the radii enclosing 50\%, 82\%, 84\% and 86\% of the stellar mass.

\begin{figure}
    %\centering
    \includegraphics[width=\linewidth]{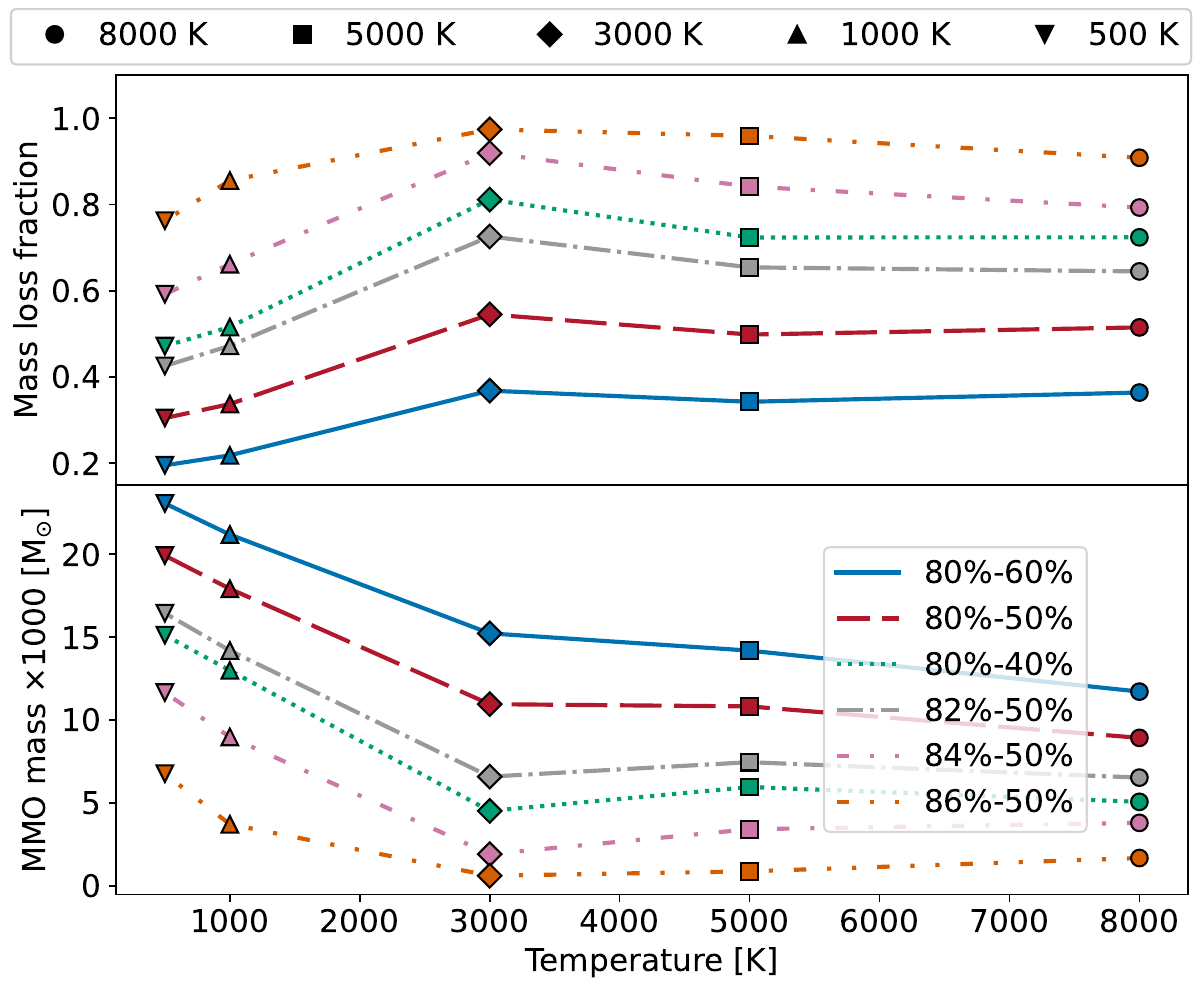}
    \caption{Final mass-loss fraction and final MMO mass for the cases $\rm M3^1_{80-40}$, $\rm M3^1_{80-50}$, $\rm M3^1_{80-60}$, $\rm M3^1_{82-50}$, $\rm M3^1_{84-50}$ and $\rm M3^1_{86-50}$ as a function of the initial gas temperature.}
    \label{fig:mass loss fraction}
\end{figure}

The radius ratio in \cref{eq:mass_loss_lombardi} has a stronger impact when the stars are more extended and is less relevant when they are more compact; thus, the mass-loss fraction estimation can be over- or underestimated depending on the adopted radii. In \autoref{fig:mass loss fraction}, we show the average collision-driven mass-loss fraction and the final MMO mass for different combinations of enclosed-mass radii. We observe a general trend of an increasing mass-loss fraction for larger differences in the enclosed-mass radii, with an average increase of $105.3\%$ and $72.5\%$ for simulations with $500~\rm K$ and $8000~\rm K$, respectively. This suggests that the collision-driven mass loss of the MMO is highly sensitive to the choice of the enclosed-mass radii. The most extended stellar configurations, corresponding to the 84\%-50\% and 86\%-50\% enclosed-mass radius combinations, yield the largest mass-loss fraction, reaching $\sim 91\%$ and $\sim 97\%$ for simulations at $3000~\rm K$. We note that even in cases where gas accretion is the dominant growth mechanism of the MMO ($T<3000~\rm K$), stellar collisions involving extended SMSs can substantially reduce the final MMO mass, with mass losses of $\gtrsim 60\%$. From the second panel in \autoref{fig:mass loss fraction}, we find that the choice of enclosed-mass radii plays a crucial role in determining whether runaway-collision scenarios can form an SMS with $M\gtrsim 10^4~\rm M_{\odot}$. Our results indicate that the formation of a VMS is always possible in the supercompetitive accretion scenario, regardless of the assumed enclosed-mass radii. In less unstable scenarios, with more massive companions ($T=8000~\rm K$), the formation of VMSs with masses of order $M\simeq 10^3~ \rm M_{\odot}$ is still possible. In contrast, SMS formation appears to occur only when the enclosed-mass radii are computed using the most compact configurations (80\%-50\% and 80\%-60\%). This suggests that under repeated stellar collisions, the formation of an SMS may be strongly suppressed in systems where stars develop highly extended envelopes. Adopting models based on stellar-evolution data or theoretical prescriptions then provides upper and lower limits on collision-driven mass loss. The corresponding final MMO efficiencies are shown in \autoref{fig:efficience}.

\begin{figure*}
    \begin{subfigure}[b]{\linewidth}
    \centering
    \includegraphics[width=0.9\linewidth]{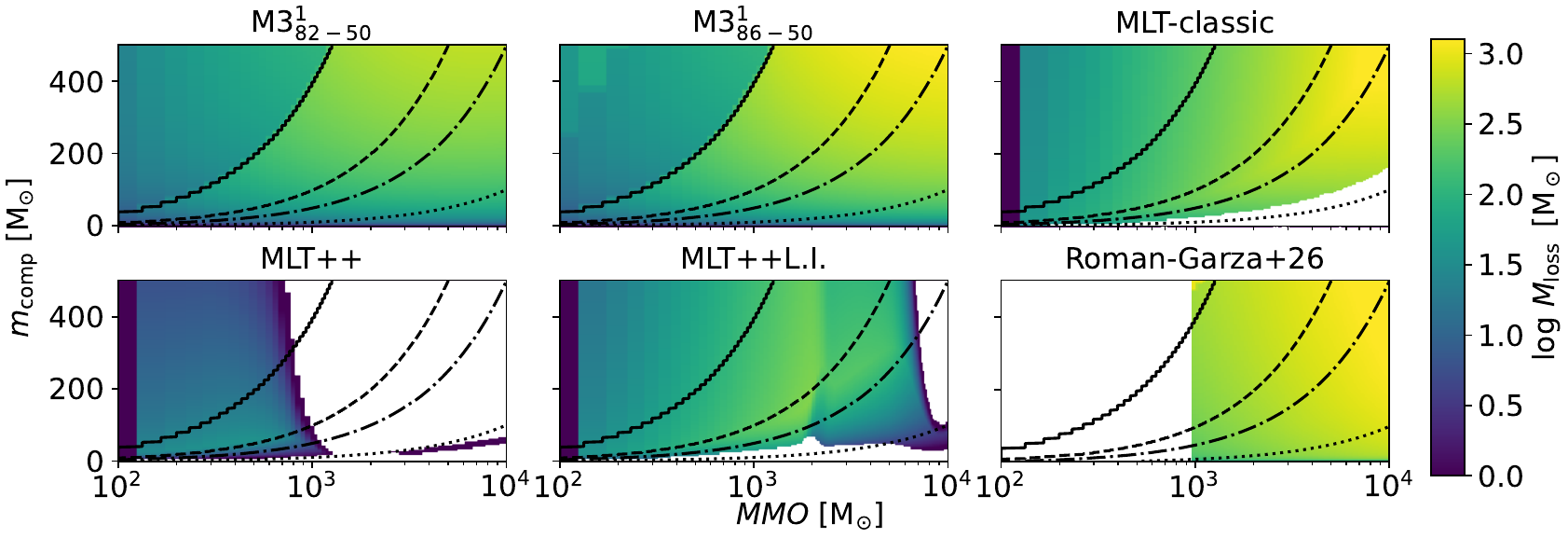}
    \caption{Absolute merger-induced mass loss.}
    \label{fig:den_mass_loss1}
    \end{subfigure}
    \hfill
    \begin{subfigure}[b]{\linewidth}
    \centering
    \includegraphics[width=0.9\linewidth]{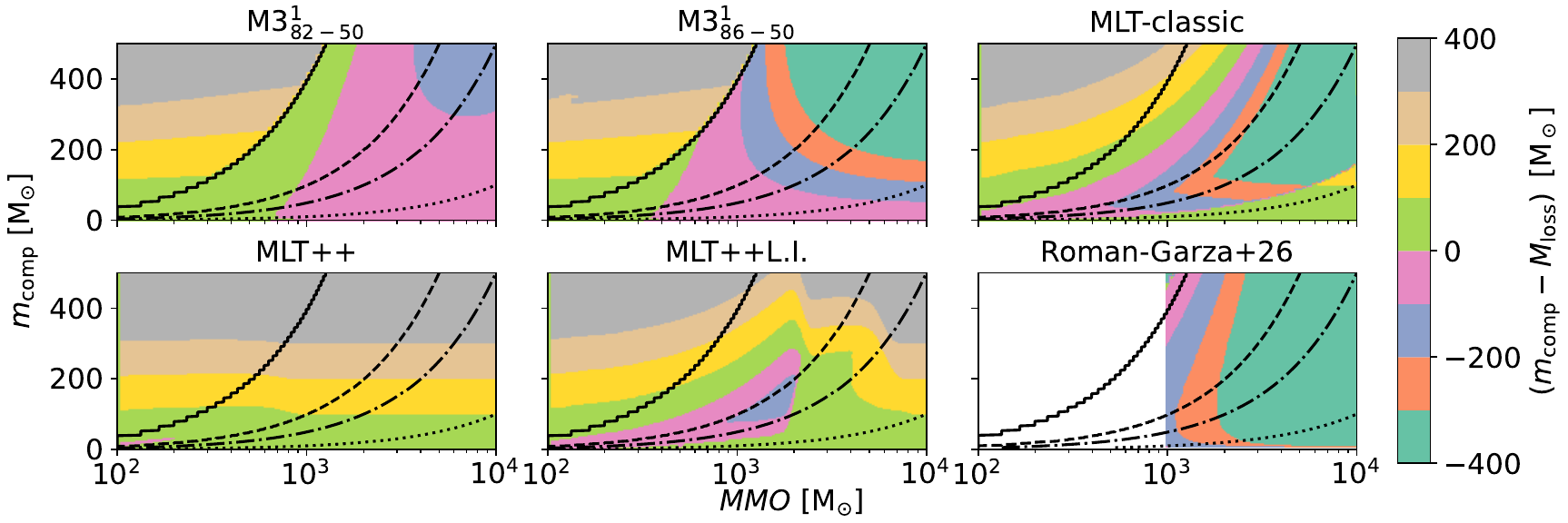}
    \caption{Net mass change $m_{\rm net} = m_{\rm comp}-M_{\rm loss}$.}
    \label{fig:den_mass_loss2}
    \end{subfigure}
    \caption{Color maps of the merger-induced mass loss (a) and net mass change (b) for models M1, $\rm M3^1_{82-50}$ and $\rm M3^1_{86-50}$, compared with the prescriptions of \citet{ramirez-galeano_collision-induced_2025} and \citet{roman-garza_massive_2026}. Because \citet{roman-garza_massive_2026} focused on collision-driven mass loss for $q<0.1$, they only explored MMO masses greater than $10^3~\rm M_{\odot}$ and companion masses from 5 to $500~\rm M_{\odot}$. The solid, dashed, dashed-dot and dotted black curves indicate the mass-ratio at $q=0.4,~0.1,~0.05$ and $0.01$.}
\end{figure*}

\begin{figure*}
    \centering
    \includegraphics[width=0.9\linewidth]{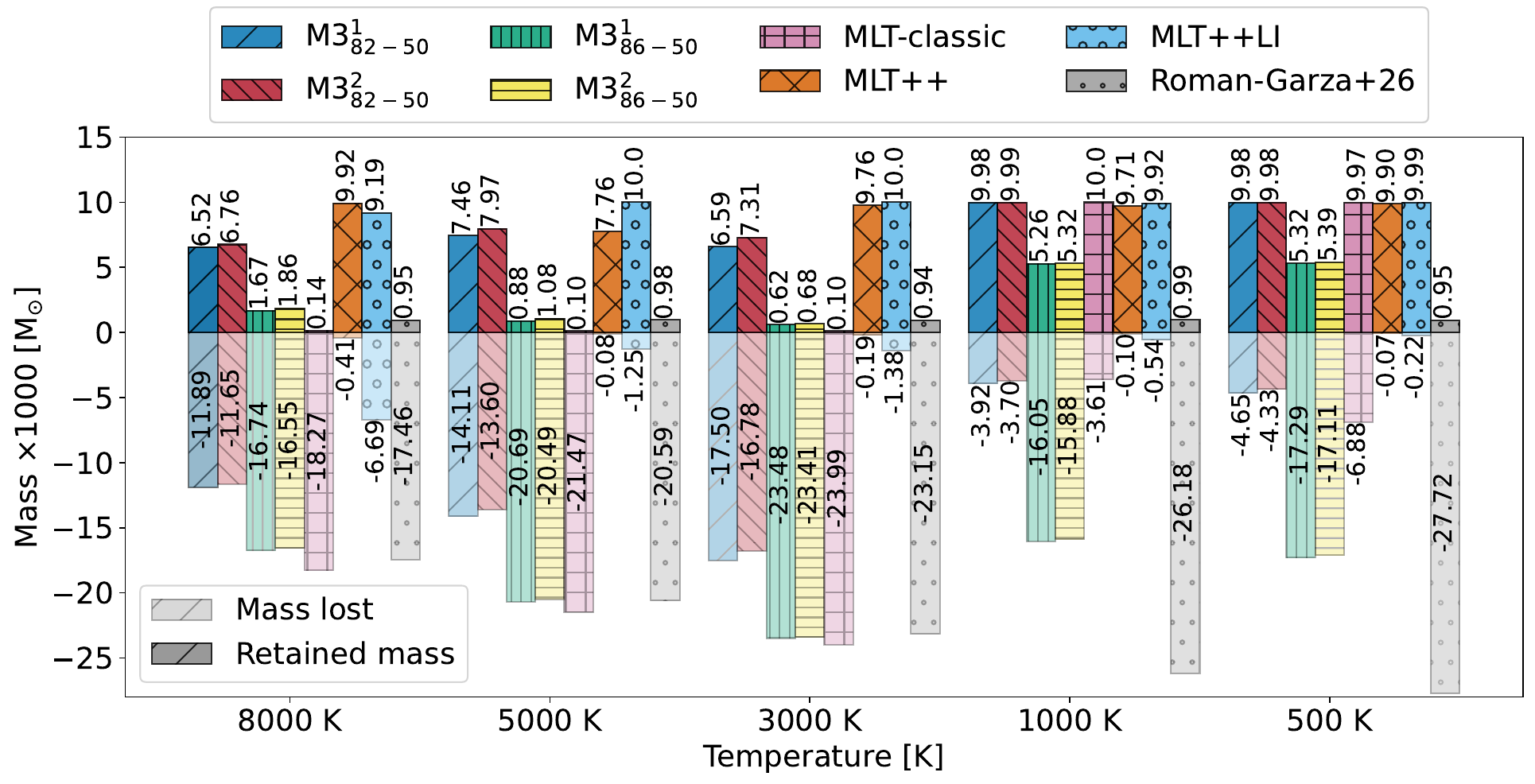}
    \caption{Final mass lost and retained by the MMO across all temperatures explored in \citet{PSolar2025}. We compare for $\rm M3^1_{82-50}$, $\rm M3^2_{82-50}$, $\rm M3^1_{86-50}$, $\rm M3^2_{86-50}$, MLT models from \citet{ramirez-galeano_collision-induced_2025} and \citet{roman-garza_massive_2026} prescriptions. General agreement for simulations with $T<1000~\rm K$ is observed for $\rm M3^{1-2}_{86-50}$ and the MLT-classic prescriptions. In systems closer to supercompetitive scenario $\rm M3^{1-2}_{86-50}$ overestimate the collision-driven mass loss in comparison with \citet{ramirez-galeano_collision-induced_2025} prescriptions.}
    \label{fig:mass_loss_teo_compa}
\end{figure*}

\section{Comparison with mass-loss estimates from the literature}

We compare our results for the models $\rm M^{1-2}_{82-50}$ and $\rm M^{1-2}_{86-50}$ with analytic mass-loss estimates by \citet{ramirez-galeano_collision-induced_2025}, who explored the impact of collision-driven mass loss and mass gain on a gas-accreting extremely massive star (EMS), with different treatments of superadiabatic convection in radiation-dominated stellar layers of EMSs. They adopted different transport prescriptions that are more efficient than classical MLT, such as the so-called MLT++ formalism, which artificially reduces the superadiabaticity in some radiation-dominated convective regions of massive stars approaching the Eddington limit, and the MLT++L.I. prescription, which artificially enhances the convective energy transport when the Eddington factor is high \citep{paxton_2011,paxton_2013,Jermyn_2023} (for more details on these MLT prescriptions, see \citet{ramirez-galeano_collision-induced_2025}).

We also compare our results with the study of \citet{roman-garza_massive_2026}, who explored merger-induced mass loss in EMSs using a hydrodynamic framework implemented within the 1-D stellar evolution code MESA \citep{paxton_2011}. They found that the ejected mass in stellar mergers is $\sim 10-30\%$. However, the published data provide only the maximum mass with positive binding energy and do not directly predict the amount of material that is ultimately ejected from the system. We therefore treat these estimates as an upper limit on the collision-driven mass loss. We use these two studies to highlight how differences in physical assumptions affect the resulting mass-loss estimates, because the dynamical configuration and the physical treatment of the merger differ between the models. In \autoref{fig:den_mass_loss1}, we compare the collision-driven mass loss predicted by models $\rm M3^1_{82-50}$ and $\rm M3^1_{86-50}$ with the estimates of \citet{ramirez-galeano_collision-induced_2025} and \citet{roman-garza_massive_2026}. The color scale shows the absolute mass loss as a function of the MMO mass and the companion mass. \citet{roman-garza_massive_2026} show what they refer to as "unbound mass", for which it is not clear whether it actually escapes and results in a real mass loss. We find clear differences among the prescriptions, where MLT++ and MLT++L.I. yield the smallest absolute mass loss, as they compute more compact structures. $\rm M3^1_{82-50}$ and $\rm M3^1_{86-50}$ predict enhanced mass loss in collisions with $q\leq 0.4$, yielding ejected masses of $0.02-637.3~\rm M_{\odot}$ and $0.02-1283~\rm M_{\odot}$. In contrast, the MLT-classic prescription predicts a more efficient regime at $q\gtrsim 0.017$, with ejected masses ranging from $12.7$ to $1557.56~\rm M_{\odot}$ (considering only collisions for which $M_{\rm loss}>0$). The most extreme losses are obtained with the prescription of \citet{roman-garza_massive_2026}, which predicts large mass loss throughout the MMO mass growth; for $q\leq0.4$, the absolute mass loss is $69.2-1547.4 ~\rm M_{\odot}$ and for $q>0.4$, it is $434.4-1107.98 ~\rm M_{\odot}$.

In \autoref{fig:mass_loss_teo_compa}, we present the final MMO masses and the corresponding collision-driven mass losses obtained with the $\rm M3^{1-2}_{82-50}$ and $\rm M3^{1-2}_{86-50}$ prescriptions, together with the predictions of \citet{roman-garza_massive_2026} and \citet{ramirez-galeano_collision-induced_2025}. For a more direct comparison, we evaluate the mass-loss fraction up to a maximum mass of $10^4~\rm M_{\odot}$. The MLT-classic and $\rm M3^{1-2}_{86-50}$ prescriptions predict larger collision-driven mass loss in simulations with more massive colliding stars than in our $\rm M3^{1-2}_{82-50}$ models, yielding an average mass loss of $\sim 14255~\rm M_{\odot}$ at $T\geq 3000~\rm K $, where the stars colliding with the MMO follow a more top-heavy mass distribution. Only in simulations with $T< 3000~\rm K$ is it possible to form an MMO with a mass of $\sim 10^4~\rm M_{\odot}$ using the MLT-classic prescription, and of $\sim 5\times10^3~\rm M_{\odot}$ when adopting the $\rm M3^{1-2}_{86-50}$ models. For the $\rm M3^{1-2}_{82-50}$ models, the formation of an MMO with $M \gtrsim 6\times10^3~\rm M_{\odot}$ is always possible. The MLT++ and MLT++L.I. models produce more compact stars, which lead to lower collision-driven mass loss and allow MMO formation in all simulations. Results obtained using the prescription from \citet{roman-garza_massive_2026} show large mass loss across all simulations, with $\gtrsim 90\%$ of the MMO mass lost through collision. We note that this prescription uses MMO masses above $10^3~\rm M_{\odot}$; therefore, no mass loss is assumed below this limit. If we assume a nonzero minimum mass loss for $M_{\rm MMO}< 10^3~\rm M_{\odot}$, this prescription predicts disruptive collisions in all simulations. These predictions can be understood from \autoref{fig:den_mass_loss2}, where we show the mass-loss and mass-gain regimes. The MLT++ model is the most conservative prescription: although the MMO loses mass, it always experiences a net mass increase. The MLT++L.I. model predicts a mass-loss regime only over a limited parameter space, namely for $M_{\rm MMO}\lesssim 10^4~\rm M_{\odot}$ and $M_{\rm comp}\lesssim 300~\rm M_{\odot}$. $\rm M3^1_{82-50}$ begins losing mass later in the MMO evolution, at $\sim 700~\rm M_{\odot}$, with the effect growing gradually as the MMO mass increases. The $\rm M3^1_{86-50}$ model enters this regime earlier, at $\sim 350~\rm M_{\odot}$. By contrast, the MLT-classic prescription predicts a mass-loss regime already for an MMO mass of $\sim 105~\rm M_{\odot}$ and a companion mass of $\sim 1.35~\rm M_{\odot}$. It yields larger mass loss for higher companion masses, while the net mass-gain region expands with MMO mass for low-mass companions; this behavior could introduce a relevant threshold in star clusters with massive colliders.

The most extreme case is the prescription from \citet{roman-garza_massive_2026}, which predicts a mass-loss regime across the full range of MMO and companion masses. This implies an unlikely scenario for SMS formation through collisions, even with high accretion rates from the CMO, as in our colder simulations. We note that the one-dimensional treatment and additional simplifications in \citet{roman-garza_massive_2026} could lead to biased mass-loss estimates. By neglecting mass transfer from the companion to the EMS, they may overestimate mass loss, because mass lost from the companion along the inspiral trajectory can increase the local EMS density, increase the drag force term, and modify the mass-loss rate. At the same time, neglecting mass transfer could underestimate the surface binding energy, making the envelope appear easier to unbind. Modeling the companion as a point mass may also overestimate collision-driven mass loss; simulations that model both colliders with SPH particles show greater kinetic energy dissipation through ram pressure, yielding less unbound mass \citep[e.g.][]{benz1992,dale_collisions_2006}. Conversely, neglecting angular-momentum transfer could lead to an underestimation of the mass loss, because angular momentum can drive envelope expansion and reduce the surface binding energy. While our method cannot capture the full hydrodynamical details and dynamical response of the collision, we find that the semi-analytic prescription of \citet{glebbeek_structure_2013} yields a net mass change broadly consistent with that obtained from detailed stellar collision models such as those of \citet{ramirez-galeano_collision-induced_2025}, particularly for the models $\rm M3^{1-2}_{86-50}$.

\section{Summary and conclusion} \label{sec:conclusions}

In this work, we have quantified how collision-induced mass loss impacts the growth of a CMO and the formation of an SMBH seed in the early Universe through a post-processing analysis in which we modeled the internal stellar structure of the colliding stars using different semi-analytic and numerical prescriptions. This analysis is motivated by recent JWST abundance measurements of high-redshift galaxies, which have spurred further investigation of enrichment channels involving rapidly growing SMSs and their collision products \citep[e.g.][]{bunker_jades_2023,cameron_nitrogen_2023,yanagisawa_strong_2024}. Using detailed stellar-evolution simulations, \citet{nandal2025} found that only SMSs in the mass range $10^3-10^4~\rm M_{\odot}$ can explain the observed $\rm N/O$ ratio for GS~3073 at redshift $z=5.55$, indicating that SMS formation is more complex than a simple monotonic CMO growth, but instead may operate only within a limited mass range.

We found that the analytic mass-radius relations M1 and M2 yield similar mass-loss fractions, reaching a minimum average efficiency of $\sim 47\%$. In addition, environments with more massive colliders show higher mass-loss fractions, reaching $\sim 23\%$ when the mean collider mass is $\sim15\rm~M_{\odot}$, and only $\sim 9.7\%$ for a mean collider mass of $\sim 5\rm~M_{\odot}$. In all cases, models based on stellar-evolution data produce higher collision-driven mass loss than the analytic models because they predict a more extended internal structure, with a mass loss of $\sim 40\%$ in the most extreme case. This occurs because extended envelopes of accreting protostars are more weakly bound, leading to larger mass-loss fractions in stellar collisions. Consistent with this picture, we also find that larger differences between the enclosed-mass radii further enhance the predicted mass loss, suggesting that collision-driven mass loss is highly sensitive to the adopted internal structure. Therefore, we note that detailed modeling of stellar evolution with the 3-D dynamics of the merger is essential to obtain reliable estimates of the final MMO efficiency when collision-driven mass loss is considered. Overall, analytic models tend to underestimate the mass-loss fraction compared to models based on 1-D stellar-structure simulations.

Our results suggest that collision-driven mass loss sets an important threshold for SMS formation, particularly in scenarios dominated by runaway collisions and/or involving massive colliders (e.g., due to higher accretion rates in the cluster or a larger Jeans mass). However, we note that uncertainties in the internal structure of accreting protostars, in collision geometry and in the treatment of energetic collisions need to be resolved and systematically explored in future studies. Our models still predict the formation of a CMO with masses of order $10^3-10^4~\rm M_{\odot}$, depending on the assumed internal structure. We compare our results with analytic and 1-D stellar-evolution prescriptions for collision-driven mass loss from \citet{ramirez-galeano_collision-induced_2025} and \citet{roman-garza_massive_2026}, where the latter predicts that SMS formation is suppressed in all simulations. The \citet{ramirez-galeano_collision-induced_2025} prescription yields a CMO with $M \sim 10^4~\rm M_{\odot}$ only in accretion-dominated simulations (which involve lower-mass companions) under the classical MLT formalism. These results are directly relevant to Little Red Dots (LRDs) at high redshift, which are often interpreted as rapidly growing black holes embedded in compact, gas-rich systems \citep[e.g.][]{Greene2024,Matthee2024,Maiolino2024,Akins2025}. If an LRD phase is connected to a collision-built SMS progenitor, our results imply that collision-driven mass loss can reduce the effective seed-formation efficiency and limit the range of cluster conditions that produce sufficiently massive seeds. In particular, systems with high accretion rates may still form SMSs.

Overall, our results show that collision-driven mass loss is an important effect that should be considered when estimating the final CMO mass, particularly in dense and highly accreting protostellar clusters. Computing the internal structure of accreting protostars using both analytic and stellar-evolution-based models provides a reasonable estimate of the mass lost during stellar collisions and can be used as lower and upper limits on the collision-driven mass loss. Our results further highlight that the internal structure of rapidly accreting protostars plays a key role in determining collision outcomes and therefore in the final growth of the CMO. In particular, high accretion rates are necessary for SMS formation but simultaneously produce more extended stellar envelopes, which enhance collision-driven mass loss. Consequently, mass loss during stellar collisions may become a limiting factor for the growth of SMBH seeds formed through runaway mergers.

\begin{acknowledgements}
PS acknowledges support through ANID/Doctorado en el Extranjero convocatoria 2022 (funding number 72220198) and thanks the German Federal Ministry of Research, Technology and Space and the German federal states (http://www.nhr-verein.de/en/our-partners) for supporting this work as part of the National High-Performance Computing (NHR) joint funding program, and we gratefully acknowledge the Kultrun Astronomy Hybrid Cluster (projects Conicyt Programa de Astronom\'ia FondoQuimal QUIMAL170001, Conicyt PIA ACT172033, and Fondecyt Iniciacion 11170268) for providing HPC resources that have contributed to the research results reported in this paper. For this work the HPC-cluster Hummel-2 at University of Hamburg was used. The cluster was funded by Deutsche Forschungsgemeinschaft (DFG, German Research Foundation) – 498394658. BR acknowledges support by the European Research Council via ERC Consolidator grant KETJU (no. 818930). DRGS gratefully acknowledges support from the Alexander von Humboldt - Foundation, Bonn, Germany and thanks for funding via the ANID BASAL project FB21003. RB acknowledges support by the Deutsche Forschungsgemeinschaft (DFG, German Research Foundation) under Germany’s Excellence Strategy – EXC 2121 "Quantum Universe" – 390833306.

\end{acknowledgements}

%-------------------------------------------------------------------
\bibliographystyle{aa}
\bibliography{spiral_biblio}

%\appendix

\begin{appendix}
\section{Mass-loss fraction for different stellar structures for $\rm M3^2$ model}
\label{appendix:m3_2_model_cases}

\begin{figure}[h]
    %\centering
    \includegraphics[width=\linewidth]{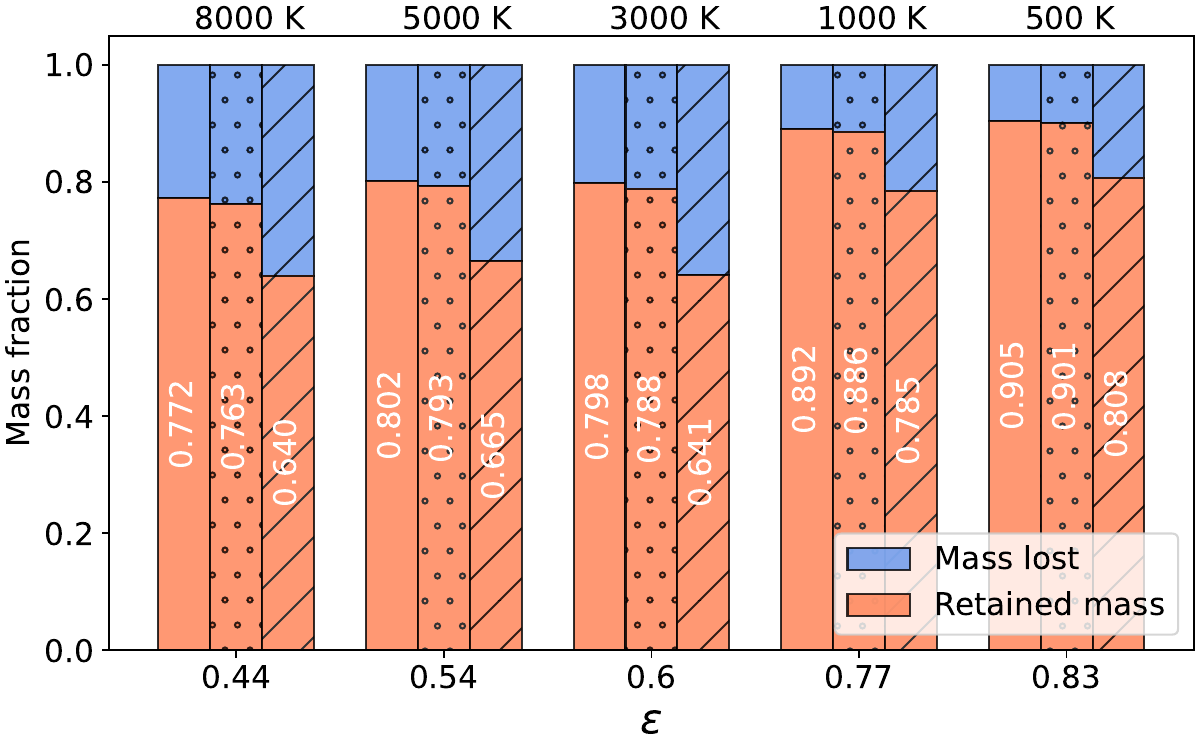}
    \caption{Average mass-loss fraction per collision and retained mass of the MMO for different initial gas temperatures as a function of the final MMO efficiency. Solid bars: mass loss with M1; dotted bars: mass loss with M2; crossed-out bars: mass loss with $\rm M3^2_{80-60}$.}
    \label{fig:mass_loss_m3_2}
\end{figure}

In this section, we explore the mass-loss fraction for different internal-structure approximations using the model $\rm M3^2$. In \autoref{table:results_model_3_2}, we summarize the collision-driven mass loss for $\rm M3^2_{80-60}$. In general, we observe a slightly smaller mass-loss fraction than for $\rm M3^1_{80-60}$. This difference does not significantly affect the final efficiency of the MMO, with a maximum difference of $\sim 1\%$. In \autoref{fig:mass_loss_m3_2}, we show the average mass-loss fraction and retained mass of the MMO for different initial gas temperatures for models M1, M2, and $\rm M3^2_{80-60}$. As in \autoref{fig:mass_loss1}, we find the general trend of larger mass loss in simulations with $T \geq 3000~\rm K$.

\begin{figure}[h]
    %\centering
    \includegraphics[width=\linewidth]{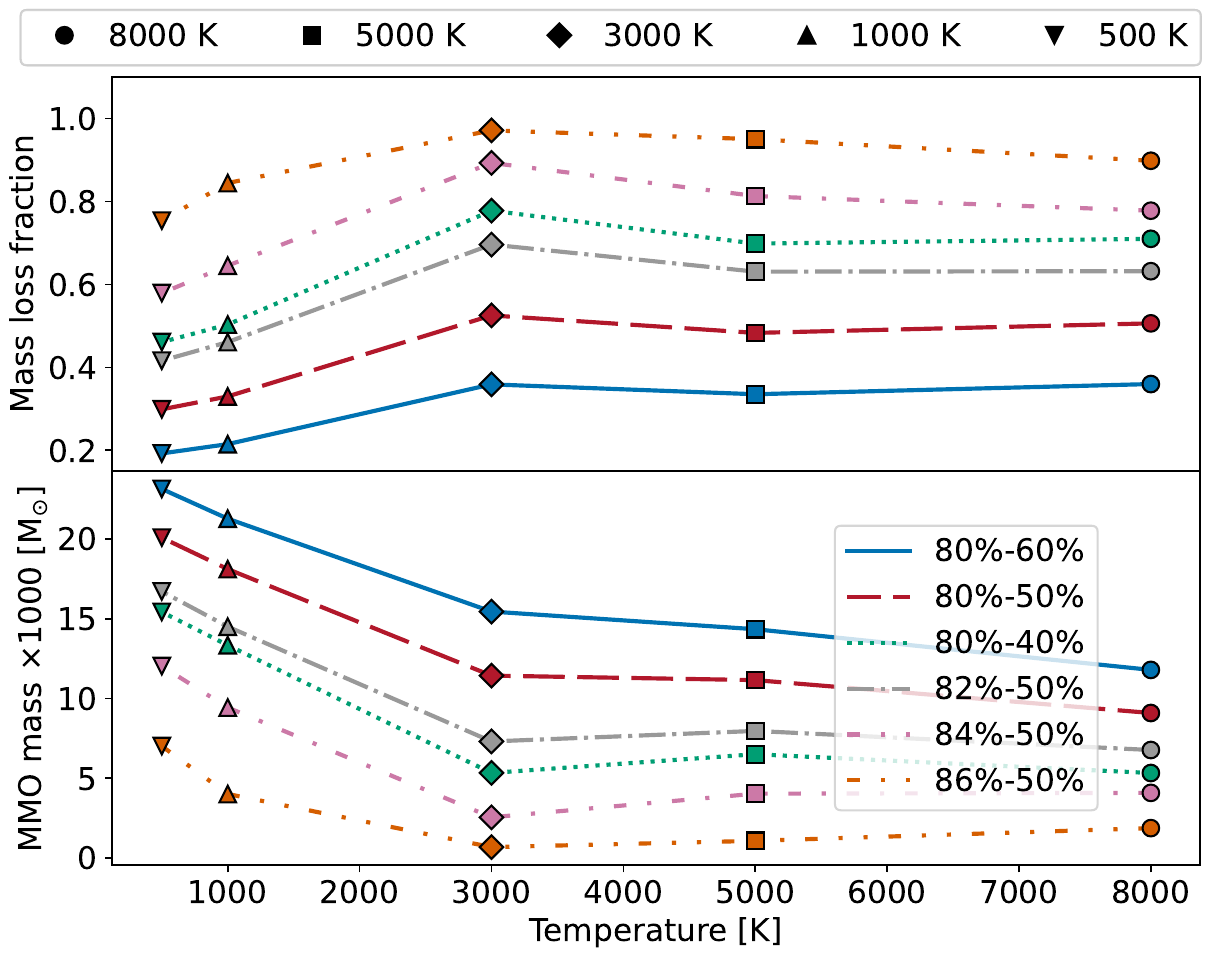}
    \caption{Final mass-loss fraction and final MMO mass for the cases $\rm M^2_{80-40}$, $\rm M^2_{80-50}$, $\rm M^2_{80-60}$, $\rm M^2_{82-50}$, $\rm M^2_{84-50}$ and $\rm M^2_{86-50}$ as a function of the initial gas temperature.}
    \label{fig:diff_m3_2}
\end{figure}

We also explore the final MMO mass with $\rm M3^2$ for different enclosed-mass radii. In \autoref{fig:diff_m3_2}, we show the average mass-loss fraction and the final MMO mass for different combinations of enclosed-mass radii. As in \autoref{subsect:stellar structure results}, we find an increase in the mass-loss fraction for larger differences in the enclosed-mass radii, with an average increase of $103.6\%$ and $71.2\%$ for simulations at $500~\rm K$ and $8000~\rm K$, respectively. $\rm M3^1$ assumes that the colliding stars are in hydrostatic equilibrium, whereas $\rm M3^2$ considers the swelling and Kelvin-Helmholtz contraction phases, predicting a less compact stellar structure during the evolution. Even so, both models predict similar mass-loss fractions across different calculations of the enclosed-mass coordinates. When stellar-evolution data are considered to model the internal structure of the colliding stars, $\rm M3^1$ predicts slightly more mass loss than model $\rm M3^2$. This occurs because $\rm M3^2$ predicts a relatively larger 50\% enclosed-mass radius than $\rm M3^1$, which further reduces the ratio $(R_{1,0.86}+R_{2,0.86})/(R_{1,0.5}+R_{2,0.5})$. As in \autoref{subsect:stellar structure results}, these results indicate that SMS formation is always possible in the supercompetitive accretion scenario, regardless of the assumed enclosed-mass radii.

 \begin{table}[h]
 \centering
 \caption{Results using $\rm M3^2_{80-60}$ prescription. }\label{table:results_model_3_2}
 \begin{tabular}{lllllll} 
 \hline 
 \hline
 Sim  & $M_\mathrm{{MMO,post}}$ & $M_\mathrm{{lost,tot}}$& $M_\mathrm{{lost,MMO}}$ &\% lost & $\epsilon_{\mathrm{post}}$ \\
  &$\times 10^3 \mathrm{[M_{\odot}]}$ & $\times 10^3\mathrm{[M_{\odot}]}$&$\times 10^3\mathrm{[M_{\odot}]}$ & &  \\
 \hline
 $1$   & 11.41 & 7.40 & 7.36 & 39.20 &0.38\\
 $2$   & 12.18 & 5.95 & 5.93 & 32.74 &0.41\\
 $3$   & 14.62 & 6.96 & 6.95 & 32.23 &0.49\\
 $4$   & 14.08 & 7.54 & 7.51 & 34.80 &0.47\\
 $5$   & 14.88 & 8.46 & 8.34 & 35.92 &0.50 \\
 $6$   & 16.03 & 9.05 & 8.96 & 35.86 &0.53\\
 $7$   & 21.02 & 4.55 & 4.55 & 17.78 &0.70\\
 $8$   & 21.54 & 7.25 & 7.25 & 25.19 &0.72\\
 $9$   & 24.62 & 3.85 & 3.85 & 13.52 &0.82\\
 $10$  & 21.67 & 7.21 & 7.21 & 24.97 &0.72\\
 \hline
 \hline
 \end{tabular}
 \tablefoot{Summary of MMO results for different temperatures using $\rm M3^2_{80-60}$ to approximate the internal structure. We report the final post-mass-loss MMO mass ($M_{\rm MMO,~post}$), the total mass lost by stellar collisions ($M_{\rm lost,~tot}$), the total mass lost by the MMO  ($M_{\rm lost,~MMO}$), the MMO mass-loss fraction ($\%$lost) and the post-mass-loss formation efficiency ($\epsilon_{\rm post}$), computed as in \cref{eq:efficiency}.}
 \end{table}

\section{\textbf{Off-axis} collisions and rotating stars} \label{subsect:head-off}

In this work, we assume that all collisions are head-on and we do not consider rotation in the collision products. However, stellar collisions are unlikely to be exactly head-on; consequently, many mergers may initially retain nonzero angular momentum, with angular velocities approaching the break-up velocity. In the literature, studies of MS, PMS and massive stellar collisions with nonzero impact parameters ($b \neq 0$) show that mass loss increases with increasing impact parameter \citep[e.g.][]{lombardi_collisions_1996,sills_evolution_2001,sills_blue_2005,ryu_magnetic_2025}. However, during subsequent evolution, the merger product may lose a large fraction of the star due to rotational instabilities and angular-momentum conservation, potentially unbinding a significant portion of its mass, and in extreme cases even disrupting it entirely. Rapid stellar rotation, especially close to the Eddington limit, can also enhance the mass-loss rate \citep{Maeder2000}. Possible channels for angular-momentum loss during post-merger evolution include disc formation and/or magnetic fields \citep[for a review, see][]{schneider_theory_2025}.

These types of encounters could be particularly relevant in the context of runaway-collision scenarios because merger products can retain a large amount of angular momentum during subsequent encounters before losing it via processes such as magnetic disc locking or magnetic braking.

\section{Energetic collisions} \label{subsect:energetic_collision}

\begin{figure}[h]
    %\centering
    \includegraphics[width=\linewidth]{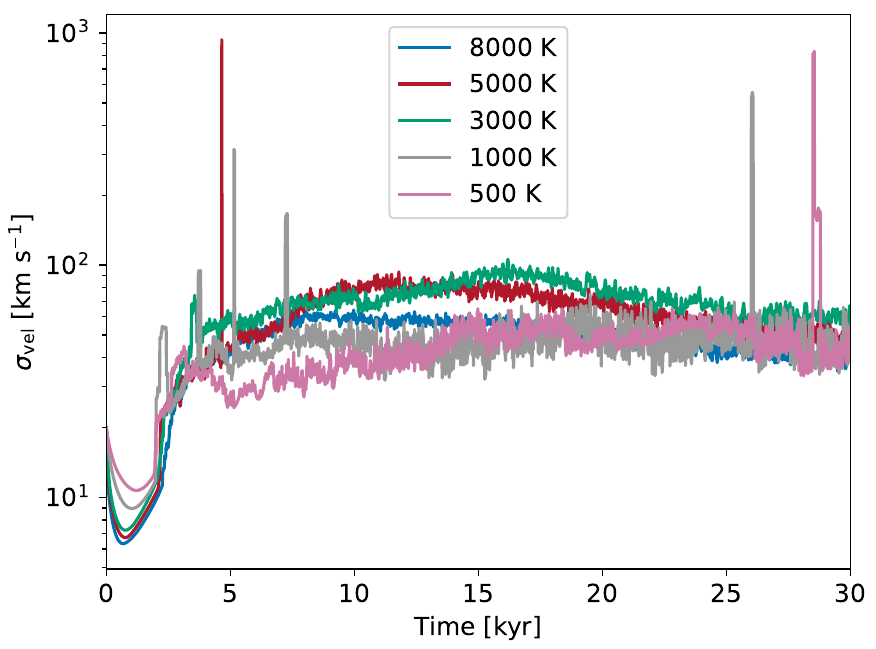}
    \caption{Time evolution of the velocity dispersion of the star cluster for all temperatures.}
    \label{fig:dispersion_velocity}
\end{figure}

\begin{figure*}[h]
    %\centering
    \includegraphics[width=\linewidth]{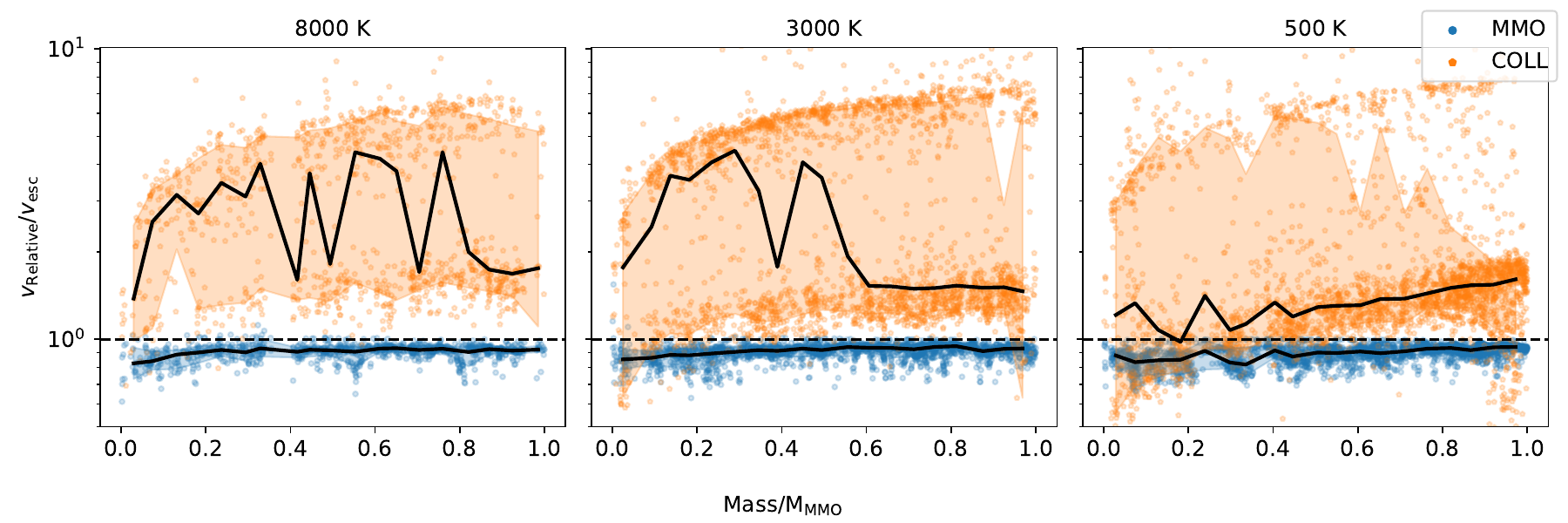}
    \caption{Ratio of the relative velocity and the stellar surface escape velocity for each collision, shown as a function of MMO mass and normalized by the final MMO mass. Orange dots represent the companion and blue dots the MMO. Black lines show the mean values, with shaded regions indicating $\pm \sigma$.}
    \label{fig:vrel-velesc}
\end{figure*}

An open question concerns the amount of mass lost in energetic stellar collisions (i.e. collisions at high relative velocity). An order-of-magnitude estimate of the collision rate in a system with a large number of particles can be obtained by computing a collision timescale, $t_{\rm coll} =\lambda/\sigma$, where $\lambda$ is the stellar mean free path and $\sigma$ is the velocity dispersion. Assuming that the stellar system is virialized, we can approximate $\sigma = (GM/R)^{1/2}$, where $M$ is the total mass and $R$ is the characteristic radius of the system. The corresponding collision timescale can be written as $t_{\rm coll} = \sqrt{R/GM(n \small{\sum_0})^2}$, where $n$ is the number density of the star cluster and $\small{\sum_0}$ is the effective cross section according to gravitational focusing. Therefore, in any virialized stellar system, the collision timescale could be approximated as $t_{\rm coll} \propto \rho^{-1}$, with $\sigma \propto \sqrt{\rho}R$. Thus, higher-density environments are expected to experience more frequent collisions and higher velocity dispersions. For typical globular clusters, $\sigma\sim 10~\rm km~s^{-1}$, but higher velocities are expected in more compact and massive clusters.

In addition, \citet{reinoso_effects_2020} analyzed the effects of a background potential in dense star clusters in the context of MBH formation through runaway collisions. They found that the background potential can increase the stellar velocity dispersion, delaying the overall cluster evolution and the formation of a central massive star. In this scenario, the SMBH seed progenitor could experience more energetic encounters, including fly-bys and potentially disruptive collisions. \citet{freitag_comprehensive_2005} studied high-velocity collisions between MS stars, with stellar masses in the range $0.1-75~\rm M_{\odot}$ and relative velocities at infinity from $0.03$ to $30$ times the stellar escape velocity. They identified different collision regimes, where at low velocities, mergers or binary formation are the most likely outcomes, with low mass-loss fractions. At higher velocities, common outcomes are fly-bys, in which the outcome is two surviving stars with significant mass loss \citep[see also ][]{benz1992,lai1993}. Disruptive collisions can also occur and are more likely at the highest velocities and smaller collision parameters. Thus, a larger relative velocity could increase the internal energy of the colliding stars and may exceed the gravitational binding energy of the remnant, leading to larger mass loss.

For our available data, we cannot calculate the relative velocity at infinity, but in \autoref{fig:dispersion_velocity} we show the velocity dispersion of the stellar cluster in the simulations for all temperatures. Simulations with colder temperatures show lower velocities after core contraction and then slowly increase to a maximum final dispersion velocity of $\sim 50 ~\rm km~s^{-1}$. The typical dispersion velocity of the cluster in a collapsing gas cloud is higher than that observed in globular clusters, which can lead to larger mass loss in collisions, implying that our estimates could be underestimated.

The effects of high-velocity collisions can be estimated by comparing the stellar surface escape velocity with the relative collision velocity. In \autoref{fig:vrel-velesc}, we present the ratio between the relative velocity and the escape velocity for both colliding stars. In most collisions, the MMO remains below but close to the mass-loss limit. This suggests that it does not lose mass in the majority of encounters. Only 3.24\%, 1.42\%, and 0.3\% of collisions in the $8000~\rm K,~3000~\rm K$ and $500~\rm K$ simulations, respectively, show the MMO above the mass-loss limit. In contrast, the collision velocity exceeds the companion's escape velocity for most encounters, implying that the companion may lose its envelope or even be fully disrupted. Our simulations may also have collisions with two surviving stars, in which both the MMO and the companion satisfy $\rm v_{relative}<v_{esc}$, because the velocity dispersion is still not high enough to disrupt the companion. In the most extreme case, this outcome occurs in 21.4\% of all collisions and is more common for less massive companions (colder simulations) and at early evolutionary stages.

\section{Collision-driven mass loss impact in Little red dots models}
\label{appendix:LRDs}

Detections by JWST of extremely compact, massive galaxies at high redshift challenge our current understanding of galaxy formation and evolution. The so-called Little Red Dots (LRDs) at redshifts $4<z<8$ have masses ranging from $10^8$ to $10^{12}~\rm M_{\odot}$ and effective radii of $3- 300~\rm  pc$ \citep[e.g.][]{Greene2024,Matthee2024,Akins2025}. Their nature remains puzzling, and several explanations for their behavior have been proposed. The presence of broad Balmer emission lines suggests that LRDs host a (super-)massive black hole at their centers with masses at least $10^7-10^8~\rm M_{\odot}$ \citep{Matthee2024,Maiolino2024}, suggesting that these black holes are overmassive relative to their host galaxies compared with extrapolations of the local black-hole-host-galaxy scaling relations. However, most LRDs are undetected in X-rays, suggesting that if MBHs are present, they must either be significantly less massive than optical estimates suggest or be obscured by extreme column densities \citep[e.g.][]{Ananna2024,Yue2024,Sacchi2025}. 

Another possible explanation is that they are intensely star-forming dusty galaxies \citep{Akins2025}. Their large stellar masses and effective radii imply that LRDs reach high stellar densities of $\sim 10^4~\rm M_{\odot}~pc^{-3}$, with extreme core densities reaching $ 10^8~\rm M_{\odot}~pc^{-3}$, making it likely that their cores enter a runaway-collision regime ($\rho_{\rm core} > 10^7~\rm M_{\odot}~pc^{-3}$), forming a CMO. Different studies have shown that SMS formation is feasible, followed by direct collapse into an IMBH through stellar collisions \citep[e.g.][]{Tagawa2020,vergara_global_2023,vergara_efficiency_2024,Fujii2024,Vergara2025b,Vergara2026}. Using semi-analytic models, \citet{Liempi2025} suggest that collision-based channels in NSCs make a relevant contribution to the total SMBH population. Recently, \citet{liempi_constraints_2026} explored whether collision-based BH channels are compatible with X-ray non-detected LRDs, finding that LRDs are ideal environments for MBH formation, particularly given the relation $R_{\rm gal} \propto M_{\rm gal}^{0.6}$ and that they should evolve into AGN, even if they were initially starburst galaxies.

The interplay and detailed evolution between gas and the stellar component remain unclear under realistic conditions in LRDs. Moderate or high accretion rates increase the stellar cross section, enhancing runaway stellar collisions. In addition, the critical mass for an NSC to form an SMBH can be reduced when gas effects are included through an external potential, which increases the velocity dispersion in the cluster and thus the collision probability \citep[e.g.][]{reinoso_effects_2020,vergara_global_2023}. However, as shown in this work, collision-driven mass loss may represent an important threshold, especially for collision-based formation channels. High dispersion velocities ($\sim 1500~\rm kms^{-1}$) could explain the broad Balmer lines; however, they could also lead to larger disruptive collisions at high mass ratios, potentially reducing or delaying SMBH seed formation.

\section{Extra figures and table} \label{appendix:extra_figures}

In this section, we show extra figures and tables of the collision-driven mass loss for the cases listed in \autoref{table:cases_resutls}.

\begin{figure}
    \includegraphics[width=\linewidth]{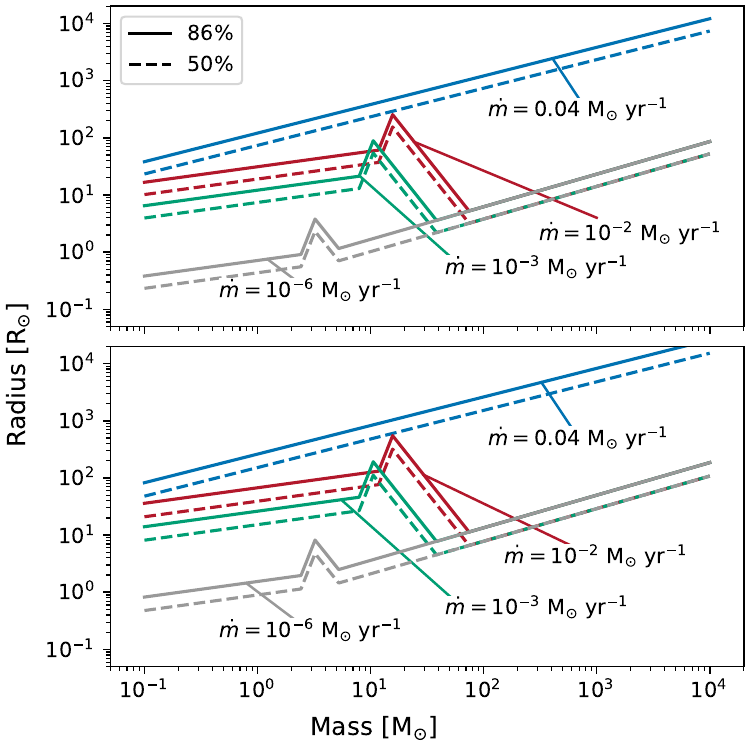}
    \caption{Mass-radius relation based on \citet{reinoso_formation_2023} (top panel; see Appendix 1 of that work for more details) and \citet{schleicher_2013} (bottom panel). The solid and dashed lines represent the radii enclosing $86\%$ and $50\%$ of the stellar mass, respectively.
    The bottom panel shows the evolutionary regimes with $t_{\rm acc}\gg t_{\rm KH}$ \pcref{eq:tacc_ll_takh} and $t_{\rm acc}\ll t_{\rm KH}$ \pcref{eq:tkh_ll_tacc} for different accretion rates.}
    \label{fig:den_86_50_appendix}
\end{figure}

\begin{figure*}
    %\centering
    \sidecaption
    \includegraphics[width=0.5\linewidth]{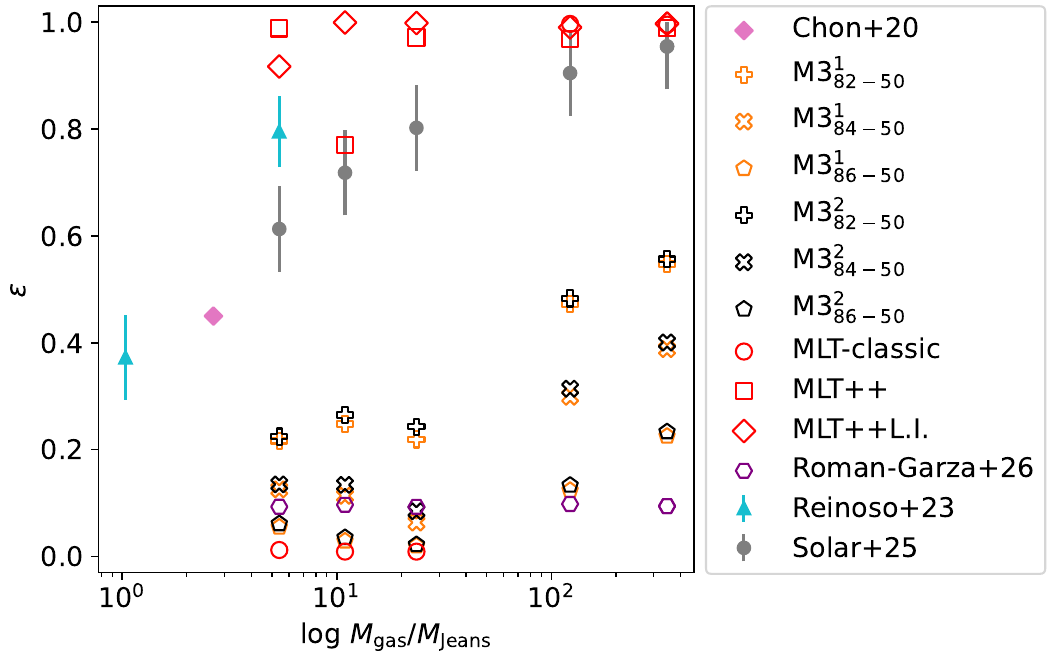}
    \caption{Efficiency of central massive object formation, considering collision-induced mass loss as a function of the gas mass divided by the thermal Jeans mass. We consider the models $\rm M3^{1-2}_{82-50}$, $\rm M3^{1-2}_{84-50}$ and $\rm M3^{1-2}_{86-50}$. We include collision-driven mass loss predictions from \citet{roman-garza_massive_2026} and different MLT prescriptions explored by \citet{ramirez-galeano_collision-induced_2025}. For these prescriptions we assume a maximum gas mass of $M_{\rm gas} = 10^4~\rm M_{\odot}$. We include data points from simulations that provide detailed models including collisions at sub-solar metallicity \citep{Chon_2020,reinoso_formation_2023,PSolar2025}.}
    \label{fig:MLT2}
\end{figure*}

\begin{figure*}
    \centering
    \includegraphics[width=0.95\linewidth]{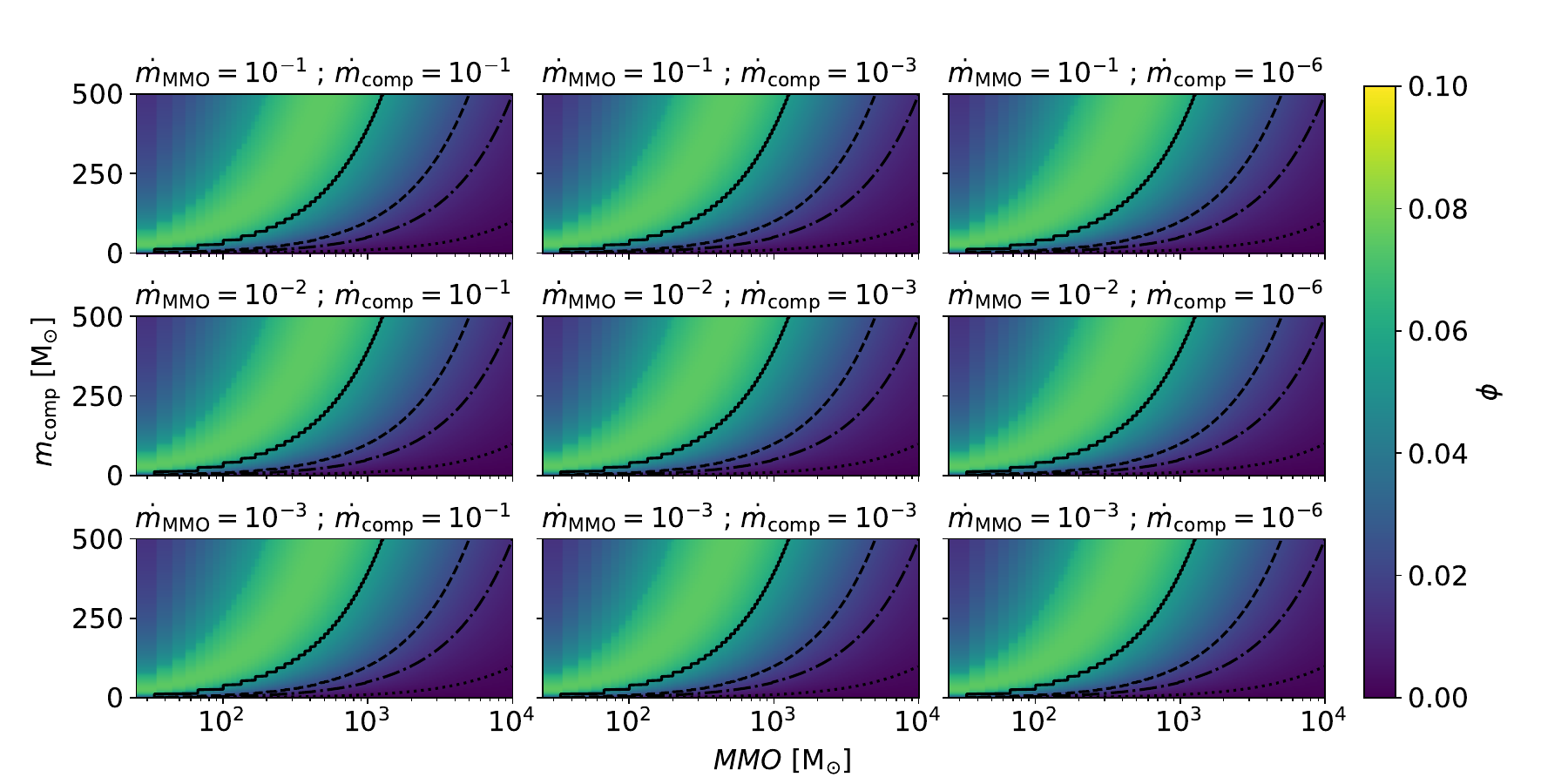}
    \caption{Color map of the mass-loss fraction as a function of the accretion rates of the most massive object (MMO) and the companion star, solving the stellar structure according to $\rm M2$. Each panel corresponds to a different combination of $\dot{m}_{\rm MMO} = 10^{-1},10^{-2}~\mathrm{and}~10^{-3}~\rm M_{\odot}~yr^{-1}$ and $\dot{m}_{\rm comp}=10^{-1},10^{-3}~\mathrm{and}~10^{-6} ~\rm M_{\odot}~yr^{-1}$. The solid, dashed, dashed-dot and dotted black curves indicate the mass-ratio at $q=0.4, ~0.1,~ 0.05$ and $0.01$.}
    \label{fig:86-50_apendix}
\end{figure*}

 \begin{table*}
 \centering
 \caption{Results of the final mass and efficiency of the MMO for different cases as we compute the internal structure.} \label{table:cases_resutls}
 %\begin{tabular}{lllllllllllll}
 \begin{tabular}{l|cc|cc|cc|cc|cc|cc} 
 \hline 
 \hline
 \multicolumn{1}{c|}{} 
& \multicolumn{2}{c|}{$\rm M3^1_{80-40}$}
& \multicolumn{2}{c|}{$\rm M3^1_{80-50}$}
& \multicolumn{2}{c|}{$\rm M3^1_{82-50}$}
& \multicolumn{2}{c|}{$\rm M3^2_{80-40}$}
& \multicolumn{2}{c|}{$\rm M3^2_{80-50}$}
& \multicolumn{2}{c}{$\rm M3^2_{82-50}$} \\
 Temp& $M_\mathrm{{MMO,post}}$ &$\epsilon_{\mathrm{post}}$ &$M_\mathrm{{MMO,post}}$ &$\epsilon_{\mathrm{post}}$&$M_\mathrm{{MMO,post}}$& $\epsilon_{\mathrm{post}}$& $M_\mathrm{{MMO,post}}$&$\epsilon_{\mathrm{post}}$& $M_\mathrm{{MMO,post}}$&$\epsilon_{\mathrm{post}}$ &  $M_\mathrm{{MMO,post}}$&$\epsilon_{\mathrm{post}}$\\
  
  $[\rm K]$&$\times 10^3\mathrm{[M_{\odot}]}$ & & $\times 10^3\mathrm{[M_{\odot}]}$ & & $\times 10^3\mathrm{[M_{\odot}]}$& & $\times 10^3\mathrm{[M_{\odot}]}$& &$\times 10^3\mathrm{[M_{\odot}]}$ & & $\times 10^3\mathrm{[M_{\odot}]}$& \\
 \hline
 $8000$&4.07 & 0.14 & 8.27 &0.28 &  5.77 &0.19 & 4.34 & 0.14 &8.44 &0.28 & 6.02 &0.20 \\
 $8000$&6.07 & 0.20 & 9.59 &0.32 & 7.29 & 0.24  &  6.32 &0.21 & 9.74 & 0.32 & 7.52 &0.25\\
 $5000$&6.40 & 0.21 & 11.16 & 0.37 & 7.78 & 0.26 & 6.98 &0.23 & 11.51 & 0.38 & 8.34 &0.28\\
 $5000$&5.53 & 0.18 & 10.49 &0.35 & 7.14 & 0.24 & 6.03 & 0.20  & 10.80 & 0.36 & 7.60 &0.25\\
 $3000$&4.93 & 0.16 & 10.73 &0.36 & 6.81 & 0.23 & 5.78 & 0.19 & 11.24 & 0.37  & 7.55 &0.25 \\
 $3000$&4.11 & 0.14 & 11.19 &0.37& 6.38 &0.21 & 4.89 & 0.16  & 11.64 & 0.39 & 7.07 &0.24 \\
 $1000$&14.91 &0.50 & 18.46 & 0.62 & 15.79 & 0.53 & 15.19 & 0.51 & 18.65 & 0.62 & 16.05 &0.53\\
 $1000$&11.09 &0.37 & 17.35 & 0.58 & 12.59 & 0.42 & 11.50 & 0.38 & 17.58 & 0.59 & 12.97 &0.43 \\
 $500$ &19.23 &0.64 & 22.41 & 0.75 & 20.16 & 0.67 & 19.46 & 0.65 & 22.53 & 0.75 & 20.36 &0.68\\
 $500$ &11.01 &0.37 & 17.43 & 0.58 & 12.73 & 0.42 & 11.40 & 0.38 & 17.65 & 0.59 & 13.08 &0.44 \\
 \hline
 \hline
 \end{tabular}
 \tablefoot{Summary of the results for different cases shown in \autoref{table:dif_stellar_radii}. We show the final post-mass-loss MMO mass ($M_{\rm MMO,~post}$) and the post-mass-loss formation efficiency ($\epsilon_{\rm post}$), computed as in \cref{eq:efficiency}.}
 \end{table*}
\end{appendix}
\end{document}